\documentclass[twocolumn,apj,numberedappendix]{aastex63}

\usepackage{chngcntr}

\usepackage{amsmath}
\usepackage{floatrow}
\usepackage{mathtools}
\usepackage{newtxtext,newtxmath}
\usepackage{amssymb}	
\usepackage{graphicx}
\usepackage{natbib}
\usepackage[all]{hypcap}								
\usepackage{hyperref}
\usepackage{xcolor}
\usepackage{braket}
\usepackage{fancyhdr}
\usepackage{bm}
\hypersetup{linkcolor=blue,colorlinks=true,filecolor=blue,citecolor=blue}

\newcommand{\hi}{H{\sc i}}

\submitjournal{APJ}

\shorttitle{CLEAN Power Spectrum}
\shortauthors{Chakraborty et al.}
\graphicspath{{./}{figures/}}

\begin{document}

\title{ A Comparative Analysis To Deal With Missing Spectral Information Caused By RFI In Cosmological \hi\ 21CM Observations}

\correspondingauthor{Arnab Chakraborty}
\email{arnab.phy.personal@gmail.com, arnab.chakraborty2@mail.mcgill.ca}

\author[0000-0002-0786-7307]{Arnab Chakraborty}
\affiliation{Discipline of Astronomy, Astrophysics and Space Engineering, Indian Institute of Technology Indore, Indore 453552, India}

\affiliation{ Department of Physics, McGill University, 3600 rue University, Montreal, QC H3A 2T8, Canada}
\affiliation{McGill Space Institute, McGill University, 3550 rue University, Montreal, QC H3A 2A7, Canada}
 
\author{Abhirup Datta}
\affiliation{Discipline of Astronomy, Astrophysics and Space Engineering, Indian Institute of Technology Indore, Indore 453552, India}

\author{Aishrila Mazumder}
\affiliation{Discipline of Astronomy, Astrophysics and Space Engineering, Indian Institute of Technology Indore, Indore 453552, India}

\begin{abstract}
We investigate the effect of radio-frequency interference (RFI) excision in estimating the cosmological \hi\ 21 cm power spectrum. Flagging of RFI-contaminated channels results in a non-uniform sampling of the instrumental bandpass response. Hence, the Fourier transformation (FT) of visibilities from frequency to delay domain contaminates the higher foreground-free delay modes, and separating the spectrally fluctuating \hi\ signal from spectrally smooth foregrounds becomes challenging. We have done a comparative analysis between two algorithms, one-dimensional CLEAN and Least Square Spectral Analysis (LSSA), which have been used widely to solve this issue in the literature.  We test these algorithms using the simulated  SKA-1 low observations in the presence of different RFI flagging scenarios. We find that in the presence of random flagging of data, both algorithms perform well and can mitigate the foreground leakage issue. But, CLEAN fails to restrict the foreground leakage in the presence of periodic and periodic plus broadband RFI flagging and gives an extra bias to the estimated power spectrum. However, LSSA can restrict the foreground leakage for these RFI flagging scenarios and gives an unbiased estimate of the \hi\ 21 cm power spectrum. We have also applied these algorithms to the upgraded GMRT  observation and found that both CLEAN and LSSA give consistent results in the presence of realistic random flagging scenarios for this observed data set. This comparative analysis demonstrates the effectiveness and robustness of these two algorithms in estimating the  \hi\ 21 cm power spectrum from the data set affected by different RFI scenarios.

 
\end{abstract}


\section{Introduction}

The  redshifted 21 cm ``spin-flip'' transition \citep{Field1958PIRE...46..240F} of the neutral hydrogen (\hi\ )  is considered as a promising tool to understand the  global phase transition of the Universe from neutral phase to ionized phase (Epoch of Reionization; EoR) over a redshift range  ($6<  \textit{z} <15$) \citep{Field1958PIRE...46..240F,Madau1997ApJ...475..429M,Furlanetto2006PhR...433..181F,Pritchard2012RPPh...75h6901P,Loeb2013fgu..book.....L,Dayal2018PhR...780....1D}. One of the promising way to study EoR is measuring  \hi\ 21 cm power spectrum along with tomographic imaging of the intergalactic medium (IGM) using low-frequency large interferometric arrays. \citep{Somnath2001JApA...22..293B,Somnath2005MNRAS.356.1519B,Morales2004ApJ...615....7M,zaldarriaga2004ApJ...608..622Z}.  Several upcoming and ongoing telescopes have been designed to study the 21 cm power spectrum, such as Donald C.Backer Precision Array to Probe the Epoch of Reionization (PAPER, \citealt{Parsons2010AJ....139.1468P}), the Low Frequency Array(LOFAR, \citealt{Harlem2013A&A...556A...2V}), the Murchison Wide -field Array (MWA, \citealt{Li2018ApJ...863..170L}), the Hydrogen Epoch of Reionization Array (HERA, \citealt{DeBoer2017PASP..129d5001D}) and the Square Kilometer Array (SKA1 LOW, \citealt{Koopmans2015aska.confE...1K}). Also a statistical detection of post-reionization (post-EoR) \hi\ 21 cm signal holds the  key to understand the large scale structure formation, source clustering and will put constrain to the cosmological parameters \citep{Somnath2003JApA...24...23B,Somnath2005MNRAS.356.1519B,Bull2015AApJ...803...21B,Wyithe2008MNRAS.383..606W}.   

However, the redshifted \hi\ signal from the EoR and post-EoR epoch is hidden behind several orders of magnitude bright radio emissions from different Galactic and extragalactic foregrounds \citep{Somnath2005MNRAS.356.1519B, zaldarriaga2004ApJ...608..622Z, Vibor2008MNRAS.389.1319J, Matteo2004MNRAS.355.1053D}. Several methods are available in the literature to estimate the \hi\ power spectrum amidst this bright sea of foregrounds. One can model the foregrounds accurately and then subtract them from the data set \citep{Yatawatta2013A&A...550A.136Y, Procopio2017PASA...34...33P}. Previous studies have shown that foreground is spectrally smooth, whereas the redshifted \hi\ 21 cm signal has spectral structure. Due to this spectral smoothness and chromatic primary beam response,  foregrounds are localized within a `wedge' shape region in two dimensional (2D) cylindrically averaged power spectrum. A traditional way is to avoid these foregrounds dominated Fourier modes inside the wedge and search for cosmological \hi\ signal in the so-called `EoR-window'. This technique is known as foreground avoidance \citep{Abhi2010ApJ...724..526D,Parsons2012AApJ...756..165P,Vedantham2012ApJ...745..176V,Morales2012ApJ...752..137M,Morales2019MNRAS.483.2207M}.    
Along with that, there are techniques to separate spectrally smooth foregrounds from rapidly fluctuating \hi\ signal without any prior modeling of foregrounds \citep{Harker2009MNRAS.397.1138H, Chapman2013MNRAS.429..165C, Mertens2018MNRAS.478.3640M}.  
However, all methods rely on the fact that the spectral and spatial properties of foregrounds are very well known. Suppose there is some oscillating feature across the frequency band,  introduced by instrumental effects or by any post-processing methods. In that case, it will pose a hindrance in isolating the redshifted \hi\ signal from the foregrounds. 

One possible cause that can introduce rapidly fluctuating components is the excision of radio frequency interference (RFI). 
In general, RFI is localized in time or frequency domain. It is usual practice to measure visibilities with high time and frequency resolution, which helps identify RFI across time and frequency. Then those RFI contaminated samples get flagged in post-processing of the data analysis. This results in a non-uniform sampling of the instrumental bandpass response, which otherwise is smooth in general. Now, each visibility is the Fourier transform of the product of the sky signal with instrumental bandpass response. Hence, RFI flagging introduces irregularities in visibilities. To estimate the power spectrum of the cosmological signal, one needs to Fourier transform the measured visibilities from frequency ($\nu$) to delay ($\eta$) domain. During this Fourier transformation, the irregularities or non-uniform sampling of the bandpass will create rapid fluctuation across the delay axis, which is proportional to the Fourier $k_{\parallel}$ modes. This can be thought of as a  convolution of visibilities in the delay domain with the delay space PSF, which is the Fourier transform of the non-uniform instrumental bandpass. The delay space PSF will have large, rapidly oscillating sidelobes, which will create a rapidly fluctuating component when convolved with visibilities in delay space. Hence, this RFI excision causes ripple across $k_{\parallel}$ axis and the `EoR-window' will become contaminated by leakage of foregrounds.      

There are several ways existing in literature to deal with this issue. PAPER analyses have used a wide-band iterative deconvolution algorithm (WIDA) to remove the foreground components in delay space and then estimate the delay space power spectrum \citep{Parsons2014ApJ...788..106P, Jacobs2015AAS...22540301J, Kerrigan2018ApJ...864..131K}.  \citet{kolopanis2019ApJ...883..133K}  used Blackman-Harris tapering window function during  Fourier transforming to delay domain to reduce foreground leakage during FFT.  \citet{Trott2016ApJ...818..139T} used least-square spectral analysis (LSSA) to perform the line-of-sight transform from frequency to delay space instead of Fourier transform to mitigate the issues related to non-uniform bandpass shape for MWA observation. \citet{Patil2017ApJ...838...65P},\citet{Gehlot2019MNRAS.488.4271G},  \citet{Mertens2020MNRAS.493.1662M} have also used LSSA to the LOFAR data set to put an upper limit on the  EoR \hi\ 21 cm power spectrum.        
\citet{Offringa2019MNRAS.484.2866O} used Gaussian weighted interpolation of missing visibility samples due to RFI flagging before any averaging of data. However, they used the same LSSA during the transformation from frequency to delay space.

\begin{table}
\caption{Details of Simulations}

\begin{tabular}{|c|c|}
\hline
\multicolumn{2}{|c|}{Simulation parameters}\\

\hline
 Telescope & SKA-1 low \\
Min baseline & 42 m  \\
Max baseline &  2.8 km  \\
Frequency & 138-146 MHz \\
Number of channels & 126 \\
Central redshift ($z$) & 9.0 \\
\hline
          & Cosmological signal: \\
          & simulated \hi\ signal cube \\
Skymodel & + \\
         & Foreground: extragalactic point \\
         & sources and diffuse foregrounds \\
         & + \\
         &  Gaussian noise \\

\hline     
\end{tabular}

\end{table}

\begin{figure*}
\centering
\begin{tabular}{cc}

\includegraphics[width=3.5in,height=3in]{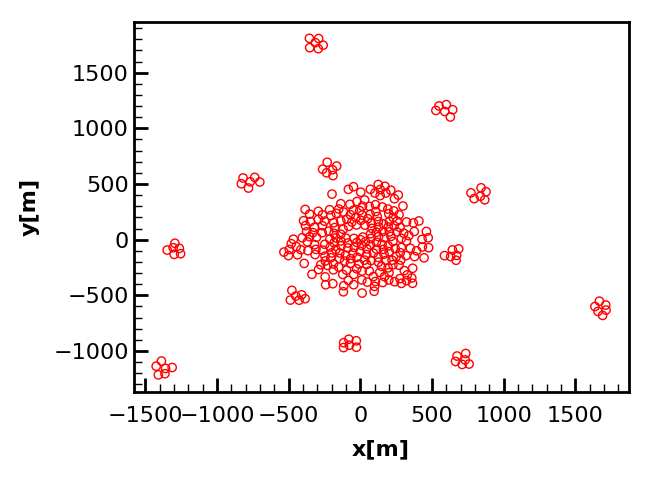} &
 \includegraphics[width=3.5in,height=3in]{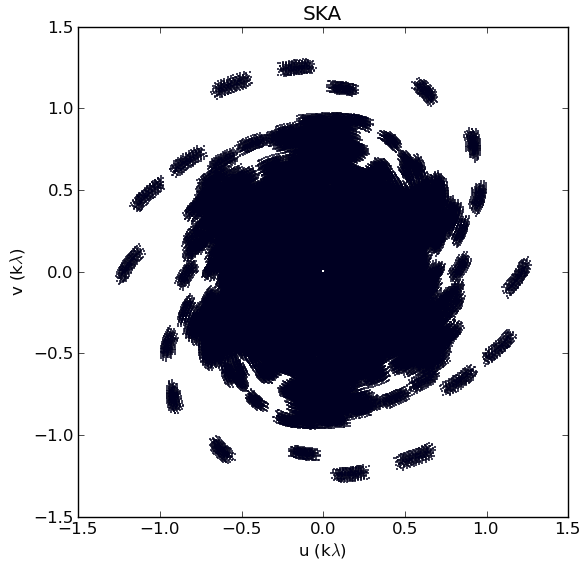} 
    \end{tabular}
    \caption{Left: the array layout of SKA-1 low with 297 elements. Right: the snapshot $uv$-coverage of the simulated data for SKA-1 low.}
    \label{ska_uv}
\end{figure*}

\citet{Deshpande1996JApA...17....7D} have used 1D CLEAN \citep{Hogbom1974A&AS...15..417H} algorithm to estimate the power spectrum of pulsar timing residual from non-uniformly sampled timing measurement. They showed that the CLEAN algorithm could help reconstruct the power spectra from non-uniformly sampled time sequences and dramatically improved the dynamic range of the estimated power spectra. \citet{Deshpande2000ApJ...543..227D} have used the same CLEAN algorithm to estimate the power spectrum of the optical depth of the cold \hi\ gas in the Galaxy toward Cas-A and Cygnus-A from the incomplete sampling of the optical depth over the extent of the source. In this work, we implement the same 1D CLEAN algorithm in the delay space to deconvolve the effects of sidelobes of the delay space PSF (point spread function). Then we reconstruct the delay spectra of visibilities and estimate the 2D and 3D power spectrum of the cosmological \hi\ signal. 

LSSA has been used in digital signal processing as well as in astronomy to deal with the non-uniformly sampled data set \citet{Barning1963BAN....17...22B, Lomb1976Ap&SS..39..447L, Stoica2009ITSP...57..843S}. The spectrum of non-uniformly sampled data is difficult to interpret as the true peak in the spectrum gives several other peaks (aliases) of various heights distributed throughout the delay spectrum. In other words, the alias structure of the major peak creates confusion in estimating the true underlying spectrum \citet{Lomb1976Ap&SS..39..447L}. This is similar to foreground contamination of large $k_{\parallel}$ modes beyond horizon limit as described before. LSSA is a commonly used method, where sinusoidal of different frequencies are fitted to the non-uniform data via the traditional least-square method, resulting in a reasonably good approximation of the spectrum of non-uniform data. The Least Square spectrum provides the best measure of the power contributed by different frequencies to the overall variance of the data and can be regarded as the extension of the Fourier methods to non-uniform data \citep{Lomb1976Ap&SS..39..447L}. In the limit of equal spacing data set, this method reduces to Fourier power spectrum. 

In this work, we show the application of these widely used CLEAN and LSSA algorithms on the simulated data (SKA-1 low) for different flagging scenarios and compare our results with that of a uniformly sampled data set (i.e., no flagging). We also apply these to the actual observation of the upgraded Giant Metrewave Radio Telescope (uGMRT) data set in the presence of realistic flagging.
This comparative analysis shows which algorithm gives a more robust and improved estimation of the cosmological \hi\ power spectrum in the presence of RFI flagging, without adding extra bias due to the missing samples issue.  

This paper is structured as follows: The simulation of data for SKA-1 low array configuration with observation setup and sky model has been discussed in Sec. \ref{Simulation}. In Sec. \ref{Formalism}, we discuss the basic formalism to estimate the power spectrum from the simulated observation, including missing samples due to flagging. The application of the CLEAN and LSSA for different flagging scenarios is shown in Sec.\ref{Application}. The application to the uGMRT observation is presented in Sec. \ref{Data}. Finally, we conclude in Sec. \ref{Colclusion}.

\begin{figure*}
\centering
\begin{tabular}{ccc}

\includegraphics[width=2.35in,height=2.25in]{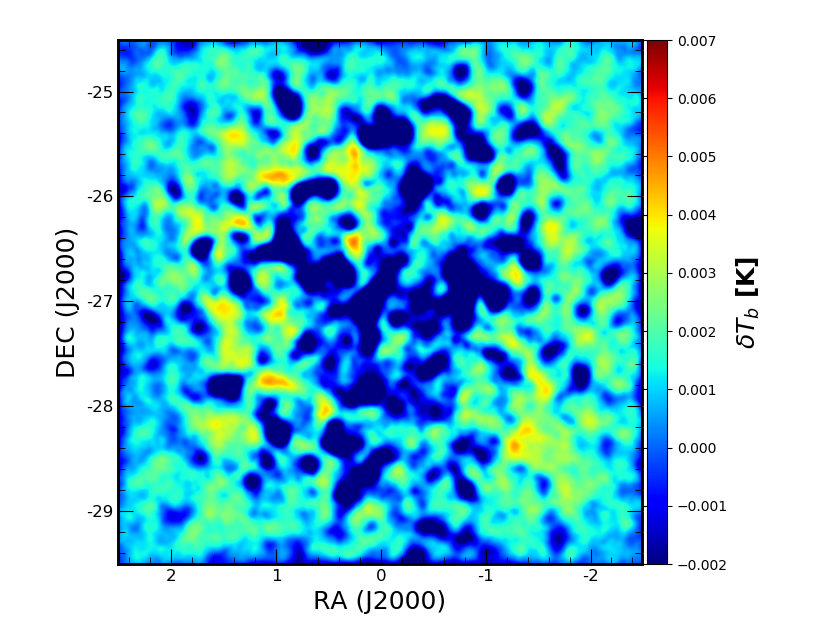} &
\includegraphics[width=2.35in,height=2.25in]{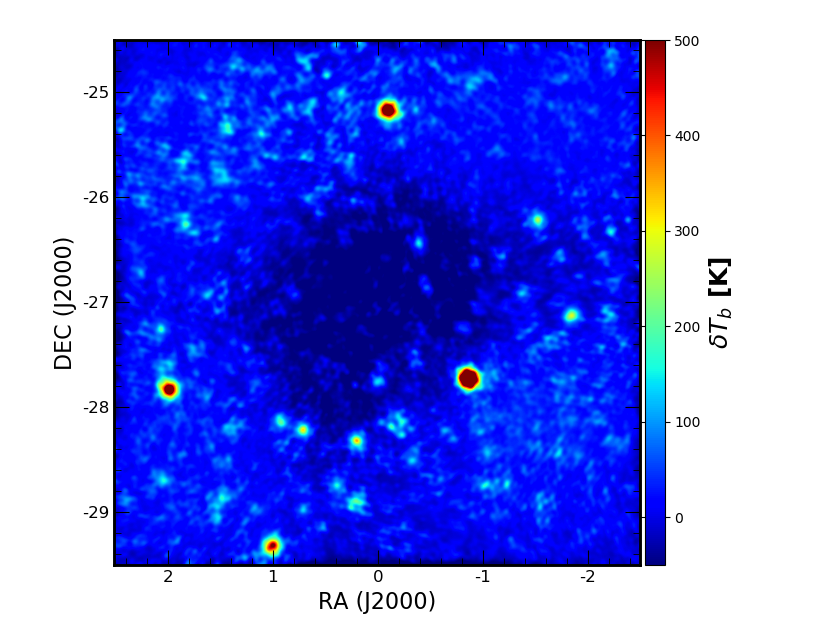} &
\includegraphics[width=2.35in,height=2.25in]{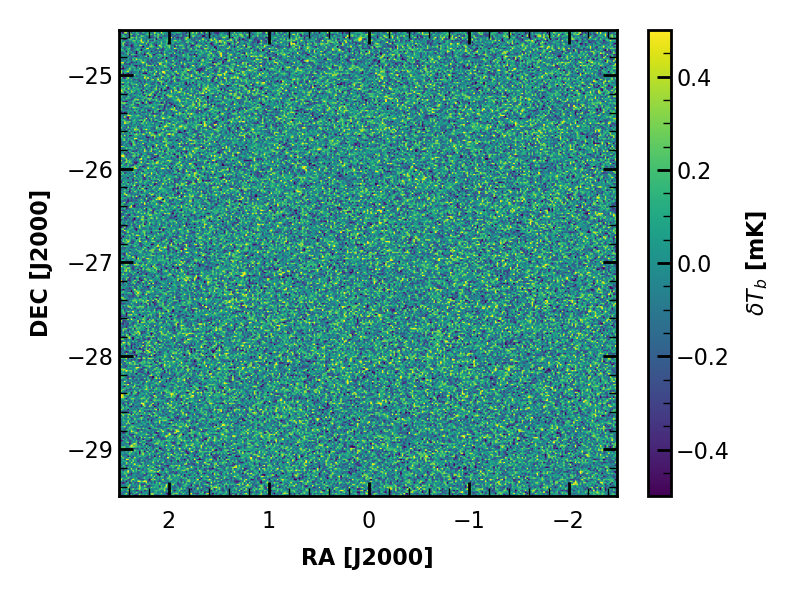}  
    \end{tabular}
    \caption{ The simulated image slices of the EoR signal (left) and foreground (middle) emissions and noise (right) at redshift $z=9$. The images cover sky areas of $5^{\circ} \times 5^{\circ}$.}
    \label{EOR_FG}
\end{figure*}

\section{Simulations}
\label{Simulation}

To investigate the effect of missing frequency channels on the estimation of the power spectrum and to demonstrate the performance of the CLEAN and LSSA, we use the data from simulated observations using the SKA-1 low. We follow the procedure given in \citet{Li2019MNRAS.485.2628L} for the simulation of EoR and foreground signal simulation and briefly mention the procedure here. \citet{Li2019MNRAS.485.2628L} extrapolate the low-frequency observation to make foreground sky maps that exhibit realistic structure of the sky. Foreground sources include the contribution from diffuse Galactic synchrotron emission, free-emission, extragalactic point sources, and radio haloes  \citep{Wang2010ApJ...723..620W, Li2019MNRAS.485.2628L}. The Galactic synchrotron emission is simulated by extrapolating the reprocessed 408 MHz Haslam map \citep{Remazeilles2015MNRAS.451.4311R}  using a power-law spectrum, the index of which is taken from \citet{Giardino2002A&A...387...82G}. The Galactic free-free radiation map has been constructed from H$\alpha$ survey map \citep{Finkbeiner2003ApJS..146..407F}, where the close relation between H$\alpha$ and free-free emission has been employed. Galactic diffuse emission is simulated at a central position of (R.A., Dec.) = $(0^{\circ}, -27^{\circ})$, which is at high Galactic latitude and suitable candidate field for EoR observation \citep{Beardsley2016ApJ...833..102B}. The extragalactic point sources are simulated by using the SKADS catalog \citep{Wilman2008MNRAS.388.1335W}. The radio halos are simulated by generating a sample of galaxy clusters with the Press-Schechter formalism \citep{Press1974ApJ...187..425P} and then using the scaling relation (cluster mass-X-ray temperature and X-ray temperature-radio power relations)  to estimate their radio emission. For more details regarding foreground simulation, please see \citep{Wang2010ApJ...723..620W, Li2019MNRAS.485.2628L}.

For the simulation of the EoR signal, we have used the {\it Evolution of 21cm Structure} \footnote{\url{http://homepage.sns.it/mesinger/EOS.html}} data set \citep{Mesinger2016MNRAS.459.2342M} and extract the image slices at corresponding frequencies from the light-cone cube of the recommended `faint galaxy' models. These EoR sky maps are tiled and re-sampled to an FoV of $5^{\circ} \times 5^{\circ}$ with a pixel resolution of $20\arcsec$. Note that this re-scaling is done to match the same sky coverage and resolution of the foreground sky maps. 

To incorporate the instrumental effects into these simulated sky maps,  we use the SKA-1 low \footnote{\url{https://astronomers.skatelescope.org/DOCUMENTS}} array configuration and  the OSKAR \footnote{\url{https://github.com/OxfordSKA/OSKAR}} package \citep{Mort2017MNRAS.465.3680M} to simulate the instrumental observation. The array configuration consists of 512 stations, each of which consists of 256 antennas randomly distributed within a circle of 35m in diameter. The stations are distributed in two parts: (1) 224 stations are distributed in a core region of diameter  1 km and the remaining distributed in three spiral arms extended up-to 35 km.  The array layout of the inner core region  is shown in Fig. \ref{ska_uv}.  OSKAR uses the Radio Interferometer Measurement Equation (RIME) of \citet{Hamaker1996A&AS..117..137H} to simulate the full stokes visibility data set, however, here we only simulate the RR component of visibilities. We provide the array configuration of SKA-1 low to the telescope model and simulate the visibilities for each  sky map at an (R.A., Dec.) = $(0^{\circ}, -27^{\circ})$.   We also include the uncorrelated Gaussian noise to our simulation corresponding to 1000 hours of observation with SKA-1 low. The system temperature ($T_{\rm sys}$) has contribution from both sky and the instrument receiver, i.e, $T_{\rm sys} = T_{\rm sky} + T_{\rm rcvr}$. The assumed sky temperature is 
$ T_{\rm sky} = 60 \rm K \Big(\frac{300 \mathrm{MHz}}{ \nu}\Big)^{2.25}$ \citep{Thompson2017isra.book.....T} and the receiver temperature is $T_{\rm rcvr} = 40 \rm K$ \citep{SKA_SWG}. We choose a 8 MHz bandwidth around the central frequency  142 MHz ($z \sim$ 9) with 64 kHz channel resolution (126 channels). The $uv$-plane corresponding to  the simulated data is being shown in Fig. \ref{ska_uv}. We then use WSCLEAN \footnote{\url{https://gitlab.com/aroffringa/wsclean}} \citep{Offringa2014MNRAS.444..606O} to make dirty image cube of size $300 \times 300 \times 126$ (pixels $\times$ pixels $\times$ $N_{\rm chans}$) for the EoR signal and foreground emission. We then  convert these cubes from Jy/beam to Kelvin (K) units \citep{Patil2017ApJ...838...65P} and then merge them to make a combined dirty image cube, $I^{D}(\theta_{x},\theta_{y},\nu)$, where $\theta_{x},\theta_{y}$ are the sky positions and $\nu$ is the frequency. We use this image cube to estimate the \hi\ 21 cm power spectrum and briefly discussed below.  The image slice corresponding to the EoR \hi\ 21 cm signal (left panel) and foreground (middle panel) and instrumental Gaussian noise (right panel)  at $z=9$ are shown in Fig. \ref{EOR_FG}.





\section{Estimation of Power Spectrum in presence of missing Frequency Channel}
\label{Formalism}
 In this section, we will briefly describe the formalism used in this work to estimate the power spectra from the dirty image cube, $I^{D}(\theta_{x},\theta_{y},\nu)$,  which also suffers from missing frequency channels, due to RFI flagging.  We follow the methodology as described in \citet{Morales2004ApJ...615....7M}, \citet{Abhi2010ApJ...724..526D}, \citet{Patil2017ApJ...838...65P}. 

 We first perform the 2D spatial Fourier transform of the image cube to make the visibility cube

\begin{equation}
    V({\bold U}, \nu) = \int I^{D}(\bm{\theta},\nu) e^{-i2\pi {\bold U} . {\bm{\theta}}} \rm d^{2}{\bm{\theta}}, 
    \label{eqn_2a}
\end{equation}
where, ${\bold U}=(u,v)$ and $\bm{\theta} = (\theta_{x},\theta_{y})$ are the vectors. 

We introduce the RFI flagging by multiplying the visibility cube with the frequency dependent sample weights ($S(\nu)$), given by:
\begin{eqnarray}
    S(\nu) &=& 0 , \forall \rm  flagged~ frequency~ channel \nonumber \\
    &=& 1, \forall \rm  channels~ without~ flagging 
\end{eqnarray}
Here $S(\nu)$ essentially corresponds to the  flagging mask due to RFI, which has been applied to the visibility data cube. 
Then we  transform the visibility cube to the delay domain ($\eta$-domain) by Fourier transforming the data along the frequency axis, given by \citep{Morales2004ApJ...615....7M}: 

\begin{equation}
    V({\bold U}, \eta) = \int V ({\bold U}, \nu) B(\nu) S(\nu) e^{-i2\pi \nu \eta} d\nu, 
    \label{eqn_2}
\end{equation}

where,  $B(\nu)$ is the instrumental bandpass response, taken flat in this case.

According to the Fourier convolution theorem, visibility $V({\bold U}, \eta)$ can be expressed as convolution of $\mathscr{F}[V({\bold U},\nu]$ with the convolving kernel $ \mathscr{F}[B(\nu) S(\nu)]$ as: 

\begin{equation}
    V({\bold U}, \eta) = \mathscr{F}[V({\bold U},\nu)] * \mathscr{F}[B(\nu) S(\nu)], 
    \label{eqn_3}
\end{equation}
where $\mathscr{F}$ denotes the Fourier transform operator and $*$ denotes the convolution. 



 We refer to the Fourier transform of  $[B(\nu) S(\nu)]$ to the delay domain as the delay space point spread function (PSF) and denote it as $D_{\rm psf}$. Since the excision of RFI contaminated channels results in a non-uniform sampling of the instrumental bandpass, which is encoded in $S(\nu)$, the simple Fourier transform of visibilities (Eqn. \ref{eqn_2}) along the frequency axis will result in spectral leakage. This causes leakage of smooth foregrounds into the EoR window. In other words, the side lobes of the delay space PSF ($D_{\rm psf}$) due to RFI flagging will be large. Convolution of $D_{\rm psf}$ with $\mathscr{F}[V({\bold U,\nu}]$ will result in leakage of foregrounds beyond the horizon limit into the EoR window and will show an oscillating pattern above the horizon delay.
 
\begin{figure}
\centering

    \includegraphics[width=\columnwidth,height=3.5in]{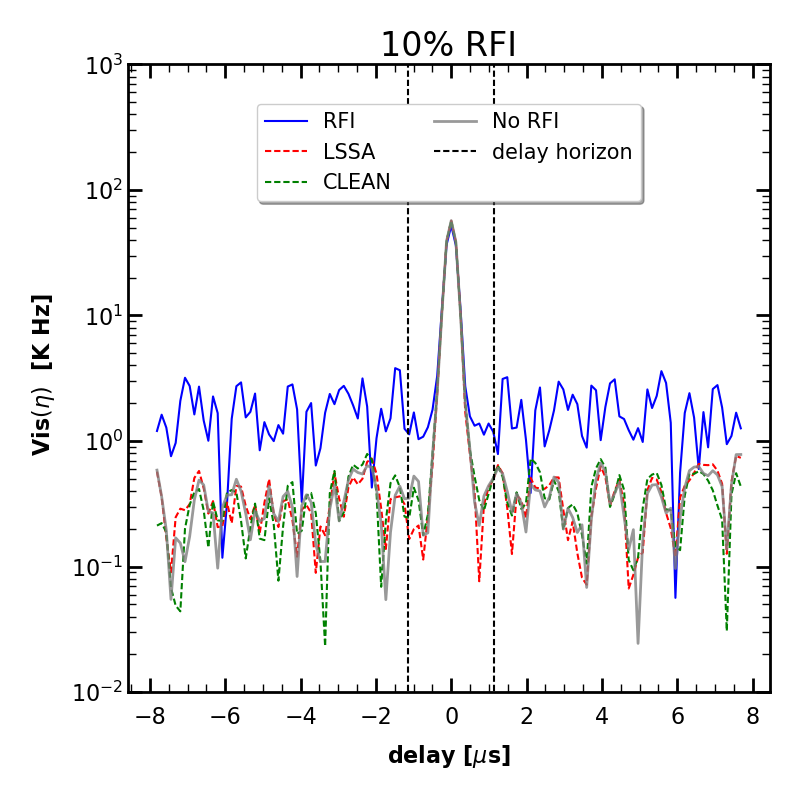} 
    \caption{ Delay spectrum of an arbitrarily chosen baseline of length 162$\lambda$.  The top blue curve is the delay spectrum with 10\% channels are randomly flagged. The green dashed curve is the delay spectrum after using 1D CLEAN algorithm across delay axis. The red dashed  curve is the delay spectrum of the baseline after using Least Square Spectral Analysis (LSSA) algorithm to this flagged data set.  The gray  curve is the delay spectrum of the baseline, when there is no flagging (no RFI). The vertical dashed black lines are the horizon limit, which is $1.14 \mu$s, for this particular baseline.}
    \label{example}
\end{figure}

 To understand the problem, we show the delay spectrum of a  single baseline  of length $162 \lambda$ (arbitrarily chosen) from our simulated data cube (see Fig. \ref{example}). We first Fourier transform the data cube to the delay domain without using any flagging, i.e, $S(\nu) = 1, \forall \rm channels$. The resulting delay spectrum of the baseline is shown in Fig. \ref{example} in gray. We refer to this as no RFI case in the plot. The vertical dashed lines are the delay horizon limits for this particular baseline. Then we randomly flagged 10\% channels and performed the delay transformation. 
 
One can use a window function $W(\nu)$ during the Fourier transformation to the delay domain to reduce the spectral leakage \citep{Thyagarajan2013ApJ...776....6T,kolopanis2019ApJ...883..133K}. Then Eqn. \ref{eqn_2} and Eqn. \ref{eqn_3} will become: 

\begin{equation}
    V({\bold U}, \eta) = \int V ({\bold U}, \nu) B(\nu) S(\nu) W(\nu) e^{i2\pi \nu \eta} d\nu, 
    \label{eqn_4}
\end{equation} and 

\begin{equation}
    V({\bold U}, \eta) = \mathscr{F}[V({\bold U,\nu}] * \mathscr{F}[B(\nu) S(\nu) W(\nu)], 
    \label{eqn_5}
\end{equation}
Here, $W(\nu)$ will act as a spectral taper during delay transformation and will reduce the spectral leakage to some extent. We show in Fig. \ref{example} the resulting delay spectrum of the baseline with a Blackman-Harris window as a spectral weighting function (in blue). As seen from the figure, there is still significant spectral leakage beyond the horizon limit even after using the window function. This is mainly due to the convolution of visibility with the  $\mathscr{F}[B(\nu) S(\nu) W(\nu)]$ , where the missing channels due to flagging is encoded inside of $S(\nu)$.  

We employ two previously widely used methods on data, CLEAN \citep{Parsons2014ApJ...788..106P} and LSSA \citep{Patil2017ApJ...838...65P, Trott2016ApJ...818..139T}, to mitigate this spectral leakage issue due to missing samples (RFI flagging). \citet{kolopanis2019ApJ...883..133K} used a Blackman-Harris tapering function to reduce the spectral leakage in their reanalysis of PAPER-64 data. However, as we have seen above, a simple spectral tapering is unable to mitigate foreground spillover due to missing samples in the presence of only 10\% RFI flagging. We briefly discuss the other two methods, CLEAN and LSSA, below and compare our results here.

{ \it \bf CLEAN:}  To restrict the spectral leakage beyond the horizon limit, we deconvolve the $D_{\rm psf}$ using a one-dimensional Hogbom CLEAN algorithm across the delay axis \citep{Hogbom1974A&AS...15..417H, Parsons2009AJ....138..219P, Parsons2014ApJ...788..106P, Kerrigan2018ApJ...864..131K}.  The CLEAN algorithm iteratively searches for the maximum peak in the delay spectrum of visibility and then subtracts a fraction (0.1; gain term) of the delay space PSF from that until a relative threshold in the residual is reached. This threshold is determined by the ratio of the initial and final amplitude in the spectrum and provided as the tolerance parameter in the CLEAN algorithm \citep{Kerrigan2018ApJ...864..131K, Ewall-Wice2021MNRAS.500.5195E}. In other words, the tolerance parameter sets the degree of foreground subtraction during CLEANing and can be set as low as we want to  CLEAN deeper \citep{Ewall-Wice2021MNRAS.500.5195E}. The choice of this tolerance parameter for various flagging scenarios is discussed in Sec. \ref{Application}. We first zero-pad the visibilities by the same number of channels as that of the simulation and apply a Tukey window (with $\alpha=0.2$) as a tapering function before performing the CLEAN \citep{Kern2019ApJ...884..105K, Ewall-Wice2021MNRAS.500.5195E}. After reaching the threshold in the residual,  we add the residual to the  CLEAN model components. This delay space  CLEANing effectively deconvolves the delay space PSF and reduces the side lobes of the delay spectrum due to non-uniform sampling of the bandpass and gives the final delay spectra. In Fig. \ref{example}, we show the delay spectrum after using 1D CLEAN across the delay axis in green dashed line. It is clear from the figure that, with this simple deconvolution technique, the leakage of foreground beyond the horizon limit due to missing channels can be restricted to minimal.

\begin{table}
\caption{Details of different cases studied in this work}

\begin{tabular}{|c|c|}
\hline
\multicolumn{2}{|c|}{Different Cases Studied in this Work}\\
\hline
Case & Flagging Scenario \\
\hline
I    & No flagging or No RFI \\
\hline
II   & 20\% random channels flagged \\
  \hline
     & 20\% random channels flagged  \\
   
 III-A &   $+$    \\ 
      &  periodic flagging, with 200 KHz flag \\
      &in every 1.28 MHz     \\ 
  \hline      
      
      & 20\% random channels flagged \\
III-B &       $+$ \\
      & periodic flagging, with 300 KHz flag \\&
      in every 0.64 MHz \\
  \hline      
      & 20\% random channels flagged \\
 &       $+$ \\
      & periodic flagging, with 200 KHz flag \\
IV      & in every 1.28 MHz \\   
      
      &  $+$ \\
      & 2 MHz wide frequency chunk from the \\ 
      & center of the band has been flagged\\

\hline                 
\end{tabular}
\label{Cases}	    
\end{table}

{\bf Least Square Spectral Analysis (LSSA):} Another widely used approach to do the line-of-sight Fourier transform of the data from frequency space to delay space in the presence of missing samples is Least Square Spectral Analysis (LSSA) \citep{Patil2017ApJ...838...65P, Trott2016ApJ...818..139T}. LSSA estimates the Fourier spectrum by fitting sinusoids to the data samples using the traditional least square method. A discrete data set is approximated by a weighted sum of sinusoids of progressively determined frequencies, using a standard linear regression or least-squares fit. If one knows the covariance of the data, then that can be used to weight the data vector to get the approximated functional form of the discrete data samples \citep{Stoica2009ITSP...57..843S, Lomb1976Ap&SS..39..447L}. \citet{Trott2016ApJ...818..139T} first used this method in their \hi\ 21 cm power spectrum estimation pipeline and applied it to the MWA data set. Later \citet{Patil2017ApJ...838...65P, Gehlot2019MNRAS.488.4271G, Mertens2020MNRAS.493.1662M} also used this methodology in their analysis of the LOFAR data set to put upper limit on \hi\ 21 cm power spectrum. We have implemented this method here as well to mitigate the spectral leakage in the delay spectrum of visibilities due to missing frequency samples.  A smooth model, which is a Fourier series up to a specified order, is being fitted to the complex visibility data cube with flags, using a linear least-square solver (scipy.optimize.lsq\_linear). We use 300 Fourier modes to fit the visibility spectrum in this analysis. The delay spectrum after application of LSSA for a single baseline is shown in red dashed line in Fig. \ref{example}. We find that similar to CLEAN, the LSSA approach is also able to reduce the spectral leakage in the presence of random RFI, and we can go down to the signal level (without RFI)  in the delay spectrum beyond the horizon limit. 

\begin{figure*}
    \centering
   
\begin{tabular}{cc}
       \includegraphics[width=3.5in,height=3.in]{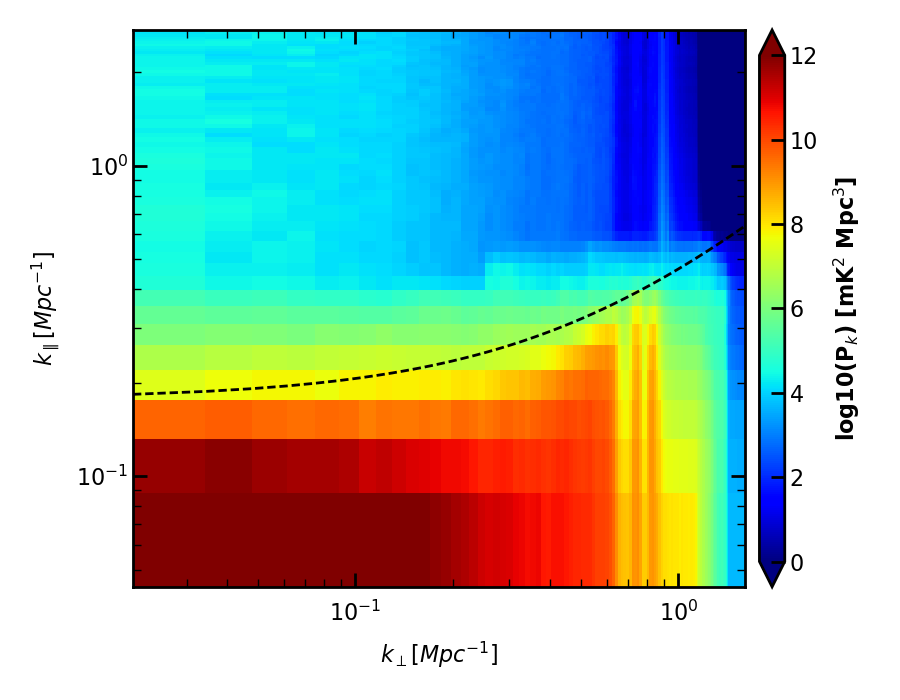} & 
        \includegraphics[width=3.5in,height=3.in]{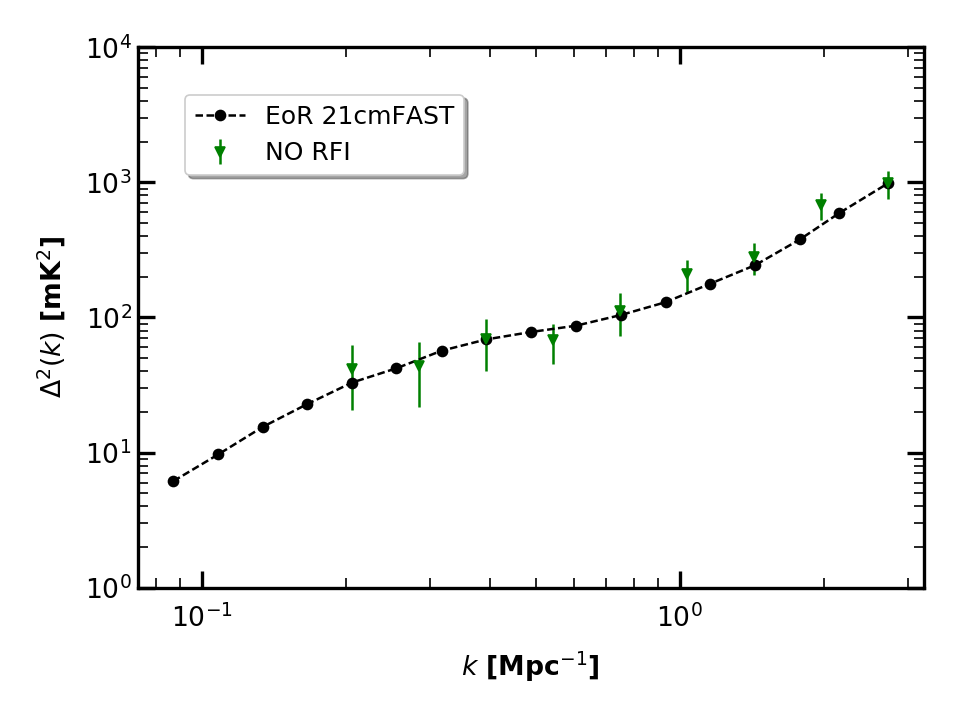} \\ 
     \end{tabular}      
         
    \caption{ Left: The cylindrically averaged 2D power spectrum in $\bold k_{\perp}$ vs $k_{\parallel}$ space for the {\bf Case I}, i.e, when there is no RFI (no flagging). The black dashed line denotes the horizon limit plus a buffer due to Blackmann-Harris window. Right: The spherically averaged 3D power spectrum using the modes above the horizon limit  with $1\sigma$ error bars (in green). We also show, in black dashed line, the input \hi\ 21 cm power spectrum of 21cmFAST.}
    \label{no_rfi}
\end{figure*}

{\bf Power spectrum estimation:} After getting the   delay visibility cube ($V (\bold U,\eta)$),  we proceed to estimate  the  cylindrically averaged 2D power spectrum using the formula \citep{Morales2004ApJ...615....7M} :

\begin{equation}
    P(k_{\perp}, k_{\parallel}) = \Big(\frac{\lambda^{2}}{2k_{B}}\Big)^{2}    \Big(\frac{X^{2}Y}{\Omega B}\Big)  |V (\bold U,\eta)|^{2}, 
    \label{ps_eqn}
\end{equation}

where $\lambda$ is the wavelength corresponding to the band-center, $K_{\rm B}$ is the Boltzmann constant,  $\Omega$ is sky-integral of the squared antenna
primary beam response, B is the bandwidth and, X and Y are the conversion
factors from angle and frequency to transverse co-moving distance (D(z))  and the co-moving depth along the line of sight, respectively (\citealt{Morales2004ApJ...615....7M}). The power spectrum $P(k_{\perp}, k_{\parallel})$ is in units of $\mathrm{K}^{2} (\mathrm{Mpc}/h)^{3}$.  The wave numbers $k_{\perp}$ and $k_{\parallel}$ are related to the baseline vector ($\bold U$) and delay ($\eta$) as : 

\begin{equation}
    k_{\perp} = \frac{2\pi |\bold U|}{D(z)},
\end{equation}{}

\begin{equation}
     k_{\parallel} = \frac{2\pi \eta \nu_{21} H_{0} E(z)}{c(1+z)^{2}},
\end{equation}

where, $\nu_{21}$ is the rest-frame frequency of the 21 cm spin flip transition of \hi\ , $z$ is the redshift to the observed frequency, $H_{0}$ is the Hubble parameter and $E(z) \equiv [\Omega_{\mathrm{M}}(1+\textit{z})^{3} +  \Omega_{\Lambda}]^{1/2}$.  $\Omega_{\mathrm{M}}$ and $\Omega_{\Lambda}$ are matter and dark energy densities, respectively \citep{Hogg1999astro.ph..5116H}. In this work, we use the best fitted cosmological parameters of the Planck 2018 analysis \citep{Planck2018arXiv180706209P} .  

The 3D power spectrum can be calculated by spherically averaging $P(\bold k_{\perp}, k_{\parallel})$ in independent $k$-bins and dimensionless 3D power spectrum can be written as \citep{Abhi2010ApJ...724..526D}:

\begin{equation}
   \Delta^{2} (k) = \frac{K^{3}}{2\pi^{2}} <P(\bold k)>_{k},  
   \label{3dps_eqn}
\end{equation}
where $k = \sqrt{k_{\perp}^{2} + k_{\parallel}^{2}}$. Throughout this work we will follow this formalism to estimate the  2D  and  3D power spectrum.  We estimate the uncertainty in the estimated power spectrum by calculating the variance of the measured power spectrum and dividing it by the number of independent modes averaged together in each $k$-bins \citep{Tegmark1997PhRvD..55.5895T}. We quote this $1\sigma$ uncertainty as the error in the 3D power spectrum at each $k$-modes.

\begin{figure*}
    \centering
   
\begin{tabular}{cccc}
       \includegraphics[width=3.5in,height=2.5in]{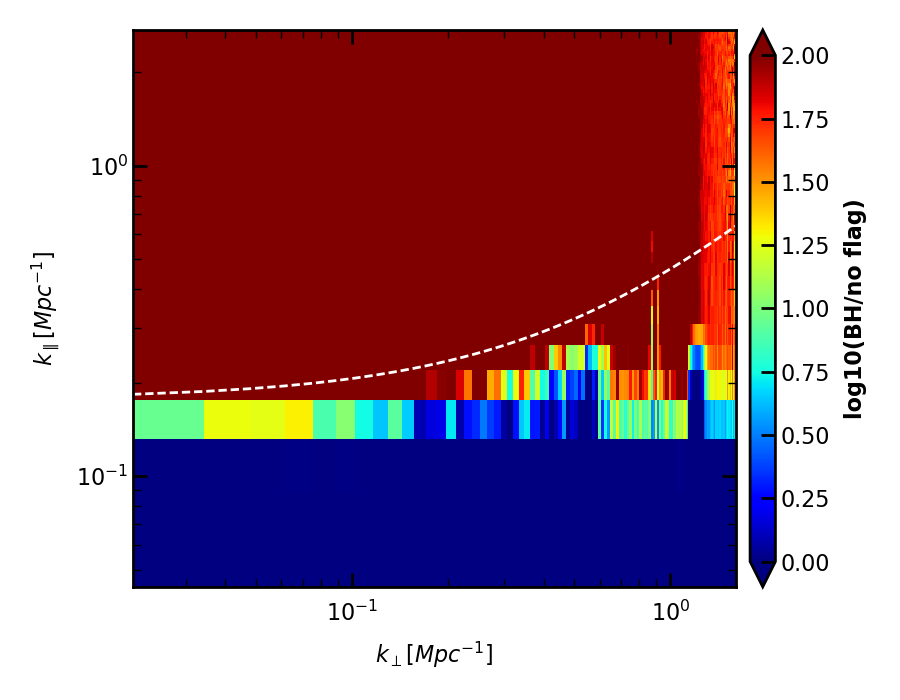} & 
        \includegraphics[width=3.5in,height=2.5in]{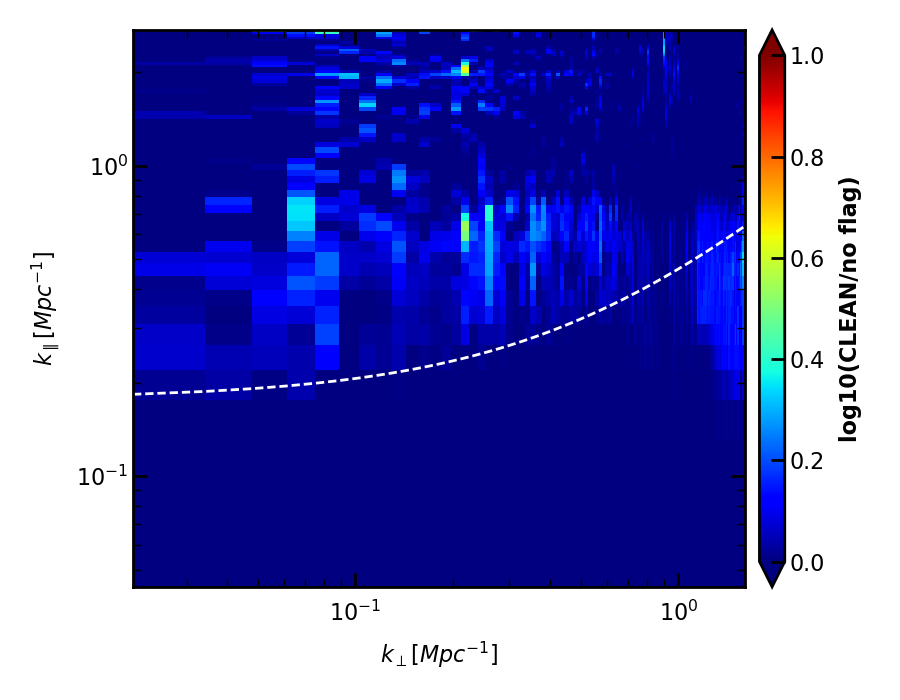} \\ 
       \includegraphics[width=3.5in,height=2.5in]{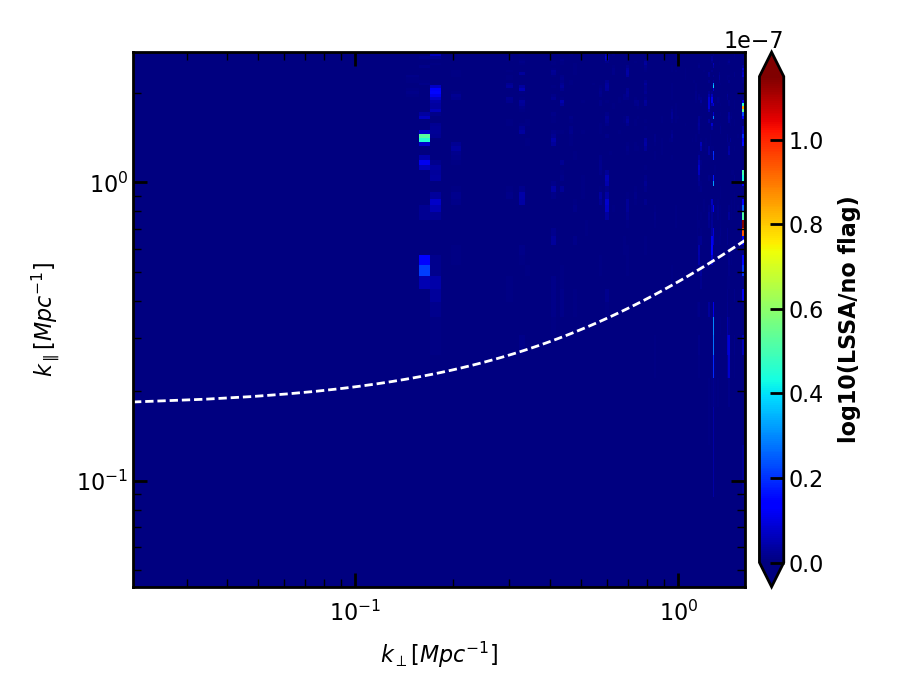} &
       \includegraphics[width=3.5in,height=2.5in]{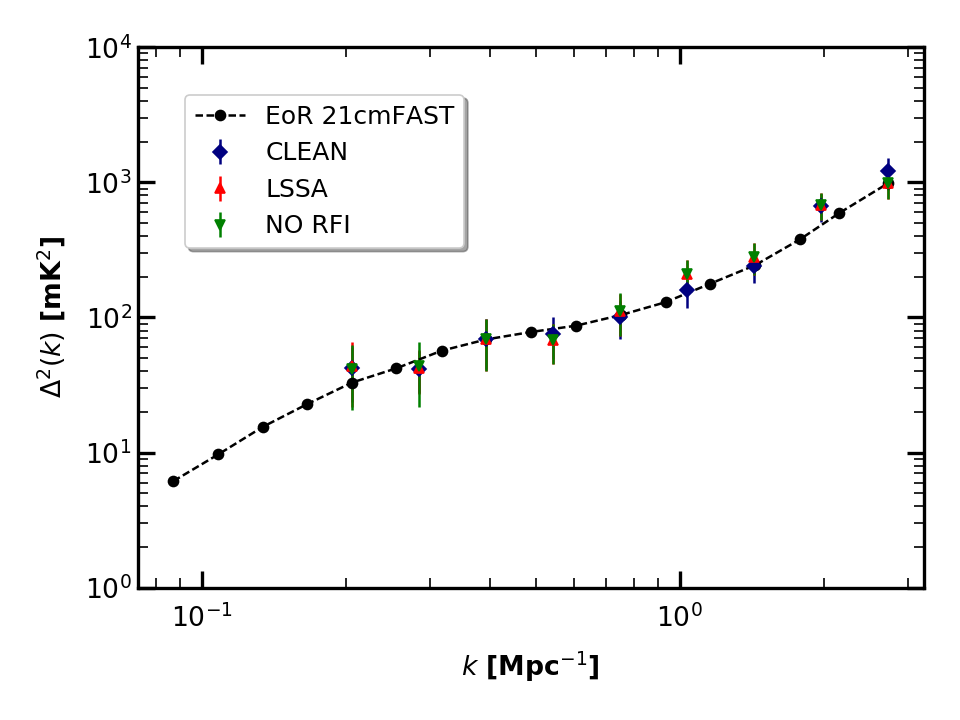}    
      
\end{tabular}      
         
    \caption{ The ratio of the cylindrically averaged 2D power spectrum between different algorithms and no RFI scenarios,  for {\bf Case II}.  Top left panel is the ratio of the flagged to no RFI scenario. Top right and bottom left panels show the ratio after applying the CLEAN and LSSA algorithms to the RFI flagged data set with respect to the no RFI scenario, respectively. 
     The bottom right panel shows the 3D power spectrum after spherically averaging the modes above the horizon limit for CLEAN (blue) and LSSA (red) algorithms. Here we also show the 3D power spectrum for the no RFI case (green) and power spectrum of the injected 21cmFAST signal (black dashed curve).}
    \label{random_flag_2d}
\end{figure*}

\section{Application on Simulated Data}
\label{Application}

To understand the effect of RFI flagging in estimating the cosmological \hi\ power spectrum, we first flagged frequency channels before performing the FT along the frequency axis using Eqn. \ref{eqn_4} and then estimated the cylindrically and spherically averaged power spectrum. Below we discuss the different cases we consider in our analysis and the corresponding results. 

\subsection{Case Studies}
To understand the effectiveness of the {\it CLEAN} and LSSA algorithms to reduce the spectral leakage for realistic flagging scenarios, we study the following cases: \\

{\bf Case I}: There is no flagging, i.e., no RFI scenario. The power spectrum is estimated from this unflagged sample. \\ 

{\bf Case II}: We randomly selected 20\% channels and flagged them.  This is the most simplistic scenario and can be observed in data taken with any radio telescope. Here, we conservatively choose 20\% random channels to flag. However, note that some of the low-frequency radio telescope s were built in radio-quiet zones, where the percentage of random flagging may be less than this. After applying the random flags to the visibility data cube, we use CLEAN and LSSA algorithms to go to the delay domain and estimate the power spectrum.  \\

{\bf Case III-A}:  The FPGA-based digital-receiver employed at MWA digitizes the beam-formed signal and performs a first stage coarse frequency channelization of the data, i.e., 1.28 MHz coarse-bands across 32 MHz bandwidth of MWA data \citep{Prabu2015ExA....39...73P}. This step applies a filter shape and aliases channels on the edge of each coarse-band (25 coarse bands across 32 MHz bandwidth) \citep{Prabu2015ExA....39...73P, Beardsley2016ApJ...833..102B}. To avoid this known aliasing effect, \citet{Dillon2015PhRvD..91l3011D} flagged the 160 KHz channel (after smoothing the data by 2 channels) in every 1.28 MHz coarse-band in their analysis of the MWA data. \citet{Beardsley2016ApJ...833..102B} also flagged two 40 KHz channels on either side of the coarse-band edges (total of four channels per coarse-band edge) to avoid the aliasing effect. In addition, they also flagged the central 40 KHz channel of each coarse-band of MWA data, which corresponds to coarse-band DC mode. These channels have been observed to contain anomalous power, likely due to very low-level rounding errors in the polyphase filter bank of the digital receivers.  \citet{Barry2019ApJ...884....1B}  also used the same flagging strategy in their updated analysis of MWA data to get a more robust upper limit on the \hi\ power spectrum. The flagging of channels at either side of each coarse-bands and at the center of the coarse-band creates a periodic flagging scenario in addition to the random flagging due to persistent RFI. To incorporate this in our analysis,  we flag 20\% randomly selected channels, and also 200 KHz flags are introduced in every 1.28 MHz across the 8 MHz bandwidth of our simulated data. We call this flagging scenario as random plus periodic flagging. \\

 {\bf Case III-B}: This is similar to the previous  {\bf Case III-A}, with the change in periodicity and increase of flagging channel width.  In this case, in addition to random flagging, we introduced 300 KHz flags in every 0.64 MHz across the 8 MHz bandwidth. \\

{\bf Case IV}: In this case, we introduce the low-level (faint) broadband RFI flagging corresponding to digital TV (DTV) signal as reported in MWA observation \citep{Wilensky2019PASP..131k4507W, Barry2019ApJ...884....1B}. \citet{Wilensky2019PASP..131k4507W} developed a pipeline, called SSINS, which can identify those ultra-faint RFI and remove the contaminated data \citep{Wilensky2019PASP..131k4507W}. \citet{Barry2019ApJ...884....1B} found a significant improvement in the estimated power spectrum after removing the low-level broadband RFI contaminated data using SSINS in their re-analysis of the MWA data set. To incorporate this broadband RFI in our analysis, we flag a 2 MHz wide frequency chunk at the center of the band, in addition to random and periodic flagging.

The different cases are also summarized in Table \ref{Cases}. We compare the results from these different cases in the following section.

\begin{figure*}
    \centering
   
\begin{tabular}{cccc}
       \includegraphics[width=3.5in,height=2.5in]{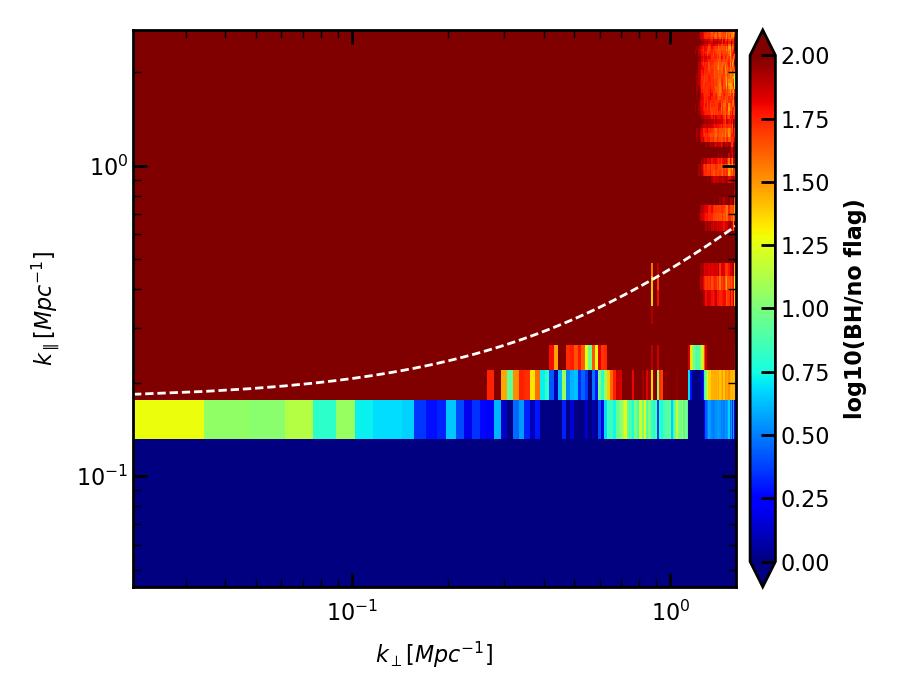} & 
        \includegraphics[width=3.5in,height=2.5in]{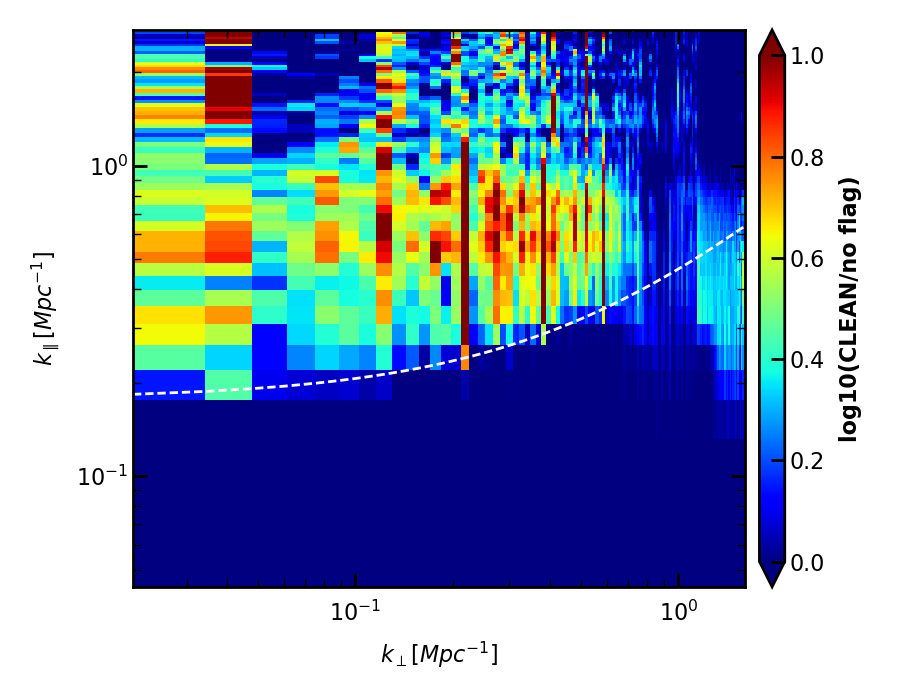} \\ 
       \includegraphics[width=3.5in,height=2.5in]{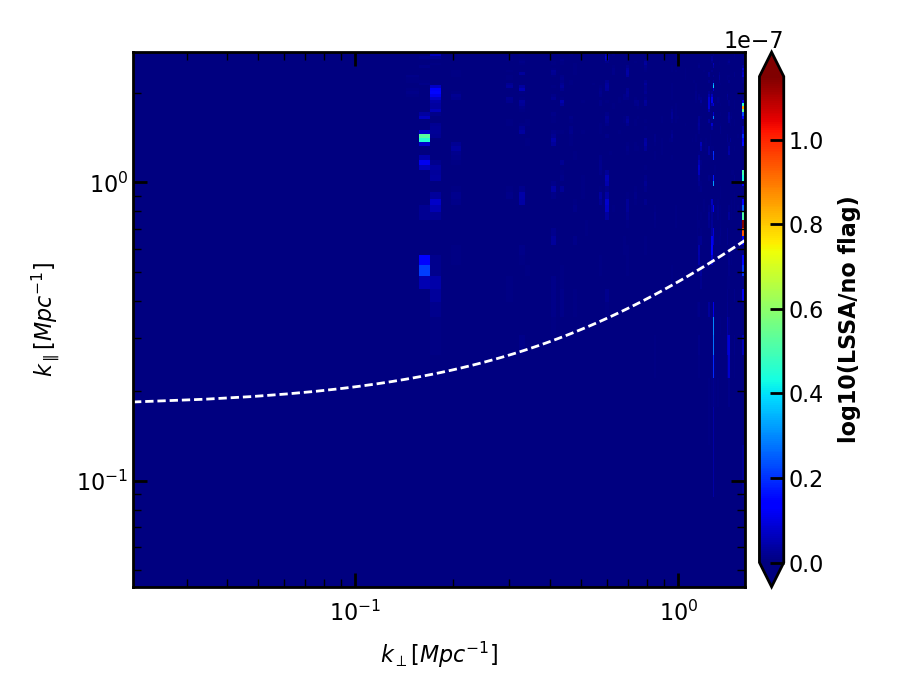} &
       \includegraphics[width=3.5in,height=2.5in]{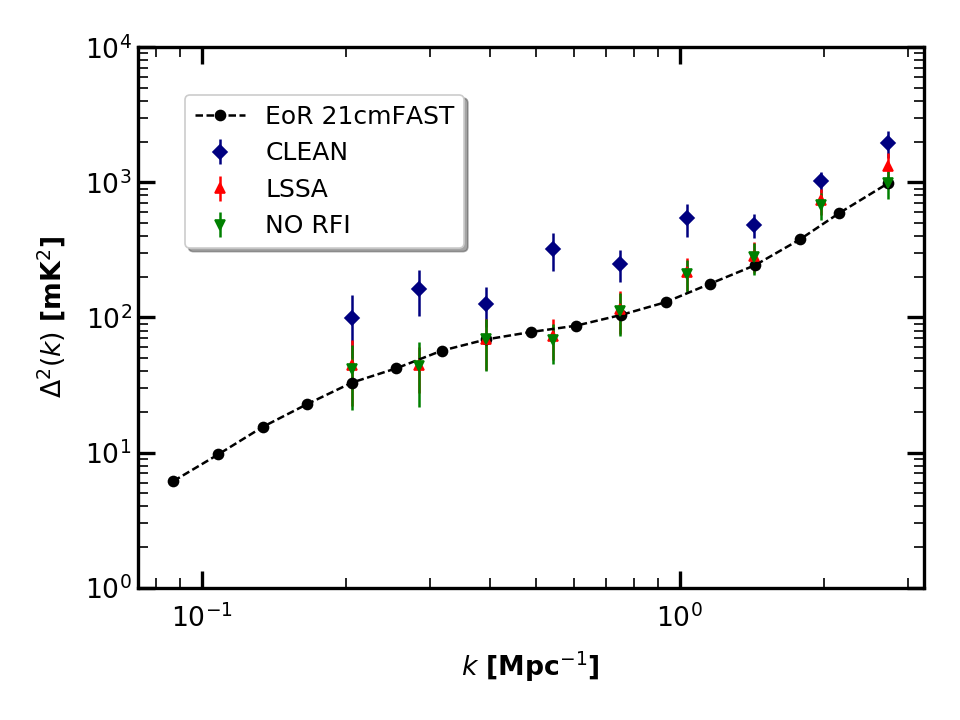}    
      
\end{tabular}      
         
    \caption{ This figure is same as Fig. \ref{random_flag_2d}, but for the {\bf Case III-A}. Here the periodic flagging is being applied to the data on top of random flagging, i.e, 200 KHz flags  in every 1.28 MHz across the 8 MHz bandwidth (see text).}
    \label{mwa_flag_2d}
\end{figure*}

   

         

\subsection{Results}
{\bf Case I}: We show in Fig. \ref{no_rfi} (left panel), the cylindrically averaged 2D power spectrum in $\bold k_{\perp}$ and $k_{\parallel}$ space (as given by Equation~\ref{ps_eqn}) when there is no RFI, i.e, no flagging involved. We find that the spectrally smooth bright foregrounds coupled with instrumental response are confined within a `wedge' shape region in this 2D space. The black dashed line is the horizon limit with an additional buffer due to the Blackman-Harris window used during FT along the frequency direction. The region above this line contains less power than the foreground wedge and is popularly known as the `EoR window'. One can avoid the foreground-dominated modes inside the `wedge' and use only the modes inside the `EoR window' to search for \hi\ 21 cm signal. This technique is called `foreground avoidance'. The spherically averaged 3D power spectrum using the modes outside the foreground wedge has been shown in the right panel of Fig. \ref{no_rfi}. The black dashed line shows the injected \hi\ 21 cm power spectrum from 21cmFAST modulated by the instrument beam. Here, we assume a simple Gaussian beam of the instrument. We find that the estimated 3D power spectrum by avoiding the foreground for no RFI scenario is consistent with the input \hi\ 21 cm signal. This case study also provides the null test, where we can recover the injected \hi\ 21 cm power spectrum using our power spectrum estimation methodology as described earlier. \\

\begin{figure*}
    \centering
   
\begin{tabular}{cccc}
       \includegraphics[width=3.5in,height=2.5in]{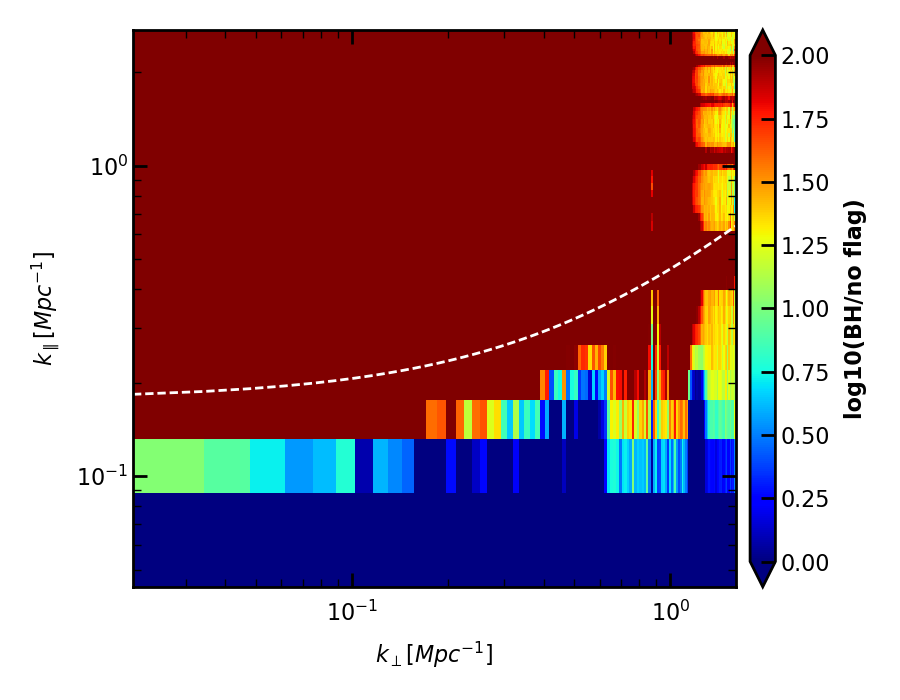} & 
        \includegraphics[width=3.5in,height=2.5in]{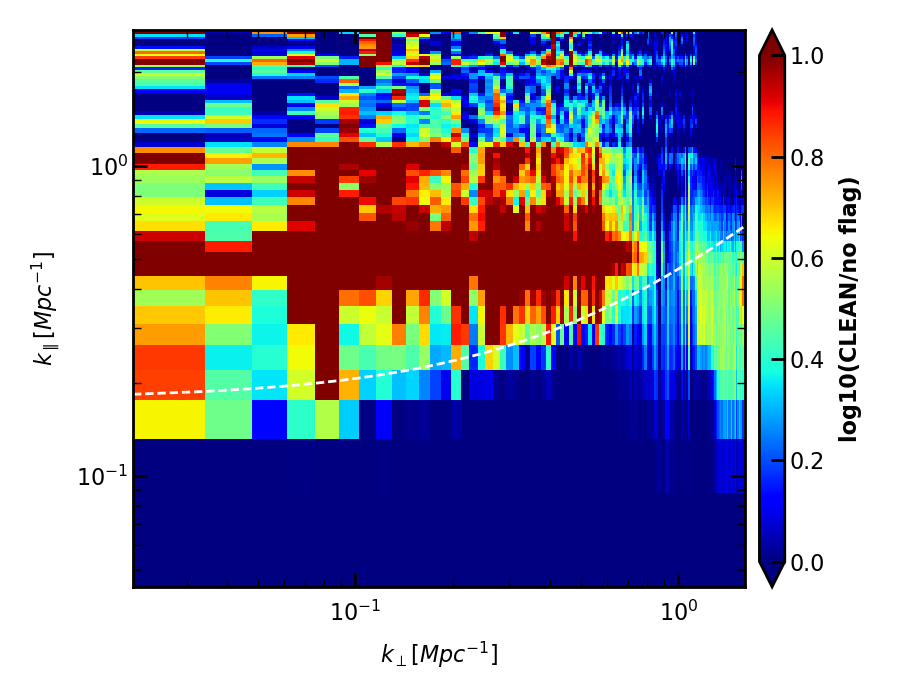} \\ 
       \includegraphics[width=3.5in,height=2.5in]{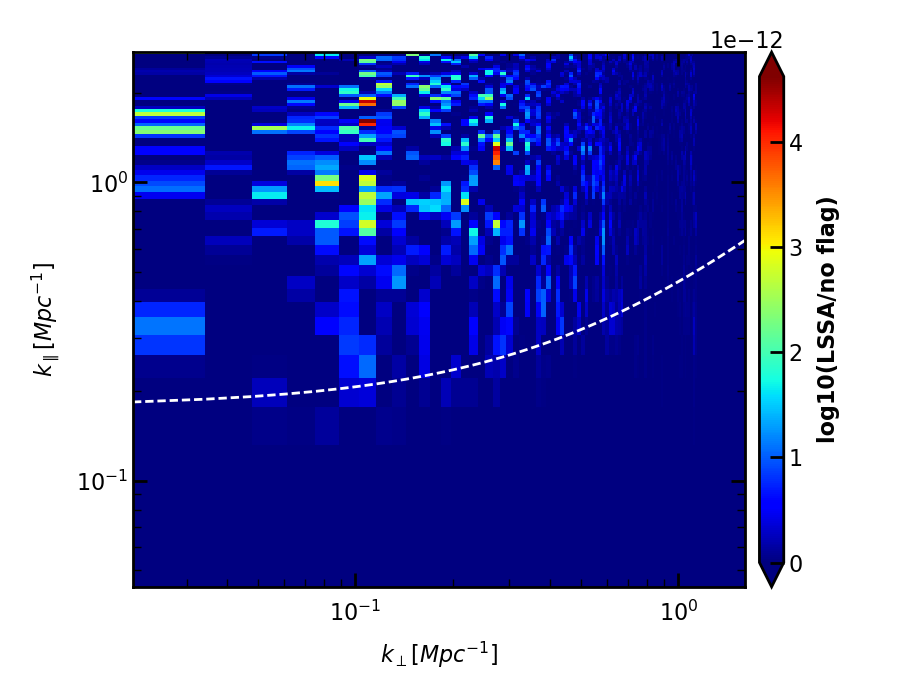} &
       \includegraphics[width=3.5in,height=2.5in]{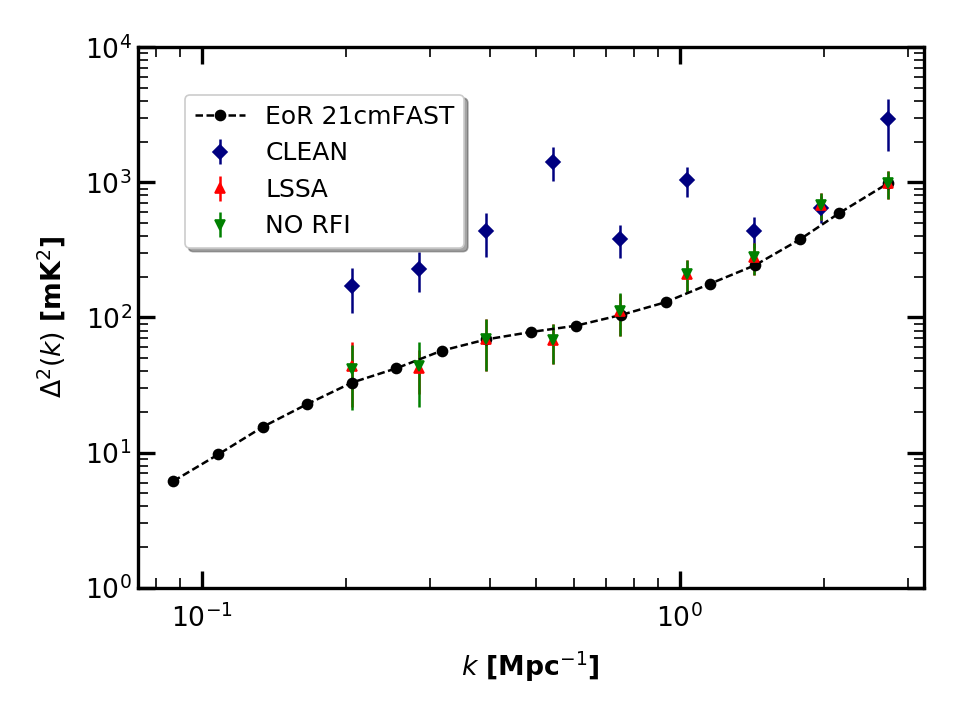}    
      
\end{tabular}      
         
    \caption{ This figure is same as Fig. \ref{mwa_flag_2d}, but for the {\bf Case III-A}, i.e, where we introduced 300 KHz flag  in every 0.64 MHz across the 8 MHz band in addition to the random flagging (see text).}
    \label{mwa_flag_2d_small_period}
\end{figure*}

{\bf Case II}: In Fig. \ref{random_flag_2d}, we show our findings for random flagging scenario. We plot the ratio of the 2D cylindrically averaged power spectrum for different algorithms applied to the randomly flagged data. At the top left of Fig. \ref{random_flag_2d}, we show the ratio of the 2D power spectrum of flagged to no RFI scenario in log-scale. We find a significant spillover of foregrounds into the `EoR window' when there are missing samples due to RFI flagging. Note that, although we use the BH window during FT along the frequency axis in this case, it completely fails to restrict the bleed-in of foregrounds into the `EoR window'.  The use of a tapering window function during the FT helps to suppress the effects of the sharp edge of the finite bandpass \citep{Thyagarajan2013ApJ...776....6T}, but it can not deal with the missing samples due to RFI flagging. At the top right and bottom left panels of Fig. \ref{random_flag_2d}, we show the ratio after applying the CLEAN and LSSA algorithms to the RFI flagged data set with respect to the no RFI scenario, respectively.  Here, we use the tolerance parameter in CLEAN to be $10^{-9}$. We find that there are only a few modes outside the foreground wedge, which have a little foreground contamination in comparison with the no RFI case when we use the CLEAN algorithm. The contamination is even less when applying LSSA to the flagged data set during FT. Nevertheless, these two plots show that one can restrict the foreground spillover to a minimum after using CLEAN and LSSA algorithms to the flagged data set in estimating the power spectrum. After applying these algorithms, one can choose a foreground-free `window' from this 2D cylindrical power spectrum and spherically average the modes to estimate the 3D power spectrum. At the bottom right of Fig. \ref{random_flag_2d}, we show the 3D power spectrum after spherically averaging all the modes above the horizon line, i.e., the modes inside the `EoR window'. We compare the result of the 3D power spectrum for CLEAN and LSSA algorithms to the no RFI scenario (Case I) and the input 21cmFAST \hi\ power spectrum. We find that we can still recover the $k-$modes of the \hi\ 21 cm power spectrum in the presence of the random flagging after using these two algorithms. \\

{\bf Case III-A}: The results for random and periodic flagging are shown in Fig. \ref{mwa_flag_2d}. Similar to the findings of {Case II}, here,  we also find that the presence of periodic flag along with the random flagging severely contaminates the `EoR window' when only the BH tapering function is being applied (top left panel of Fig. \ref{mwa_flag_2d}). At the top right panel, we show the ratio of the 2D power spectrum after the application of CLEAN to the periodic plus randomly flagged data set.   We find that the  Fourier modes inside the EoR window are still contaminated by foreground spillover after applying the CLEAN algorithm. The extra bias in CLEAN arises due to the un-subtracted foregrounds in the residual, which still contains side-lobes from RFI flagging \citep{Ewall-Wice2021MNRAS.500.5195E}. To CLEAN the spectrum much deeper than {Case II},  we choose the tolerance parameter to be $10^{-11}$. But, even after lowering the tolerance parameter, we find significant bias due to foreground leakage into the EoR window.  However, when we apply the LSSA to the flagged data set (bottom left panel), we can restrict the foreground contamination of Fourier modes beyond the `wedge' in the presence of periodic flagging. The 3D power spectrum at the bottom right panel also shows that there is an extra bias to the power spectrum for the CLEAN algorithm with respect to the no RFI scenario. But, LSSA can mitigate this foreground leakage in the presence of periodic flagging and gives an unbiased estimate of the 3D power spectrum. \\

\begin{figure*}
    \centering
   
\begin{tabular}{cccc}
       \includegraphics[width=3.5in,height=2.5in]{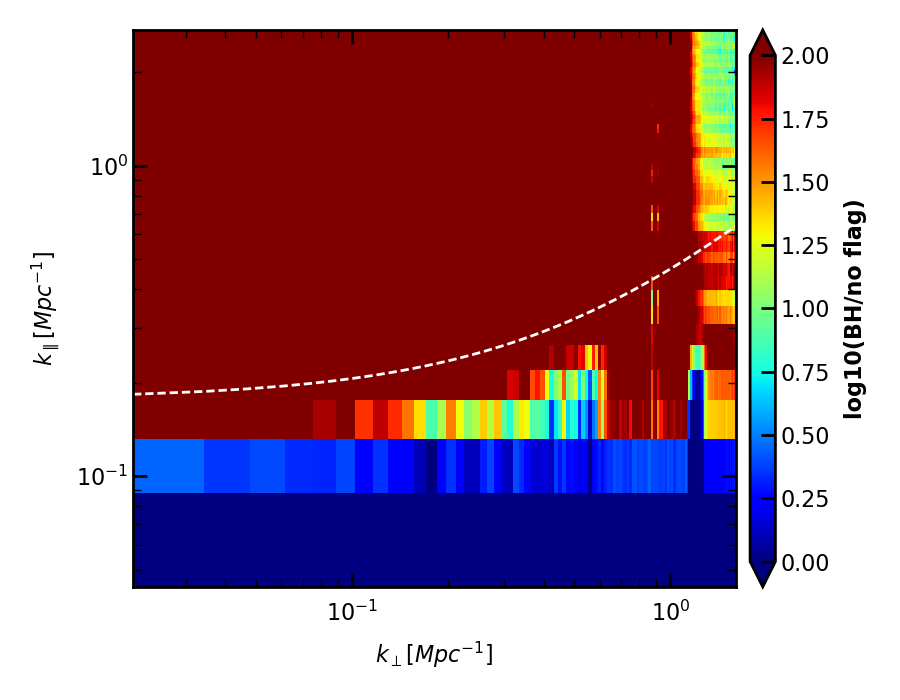} & 
        \includegraphics[width=3.5in,height=2.5in]{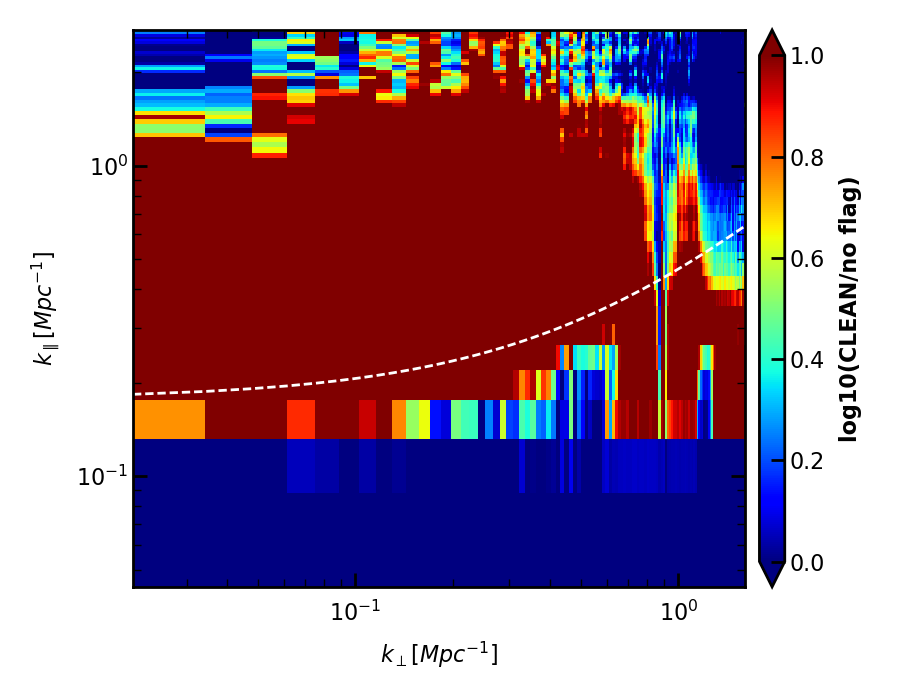} \\ 
       \includegraphics[width=3.5in,height=2.5in]{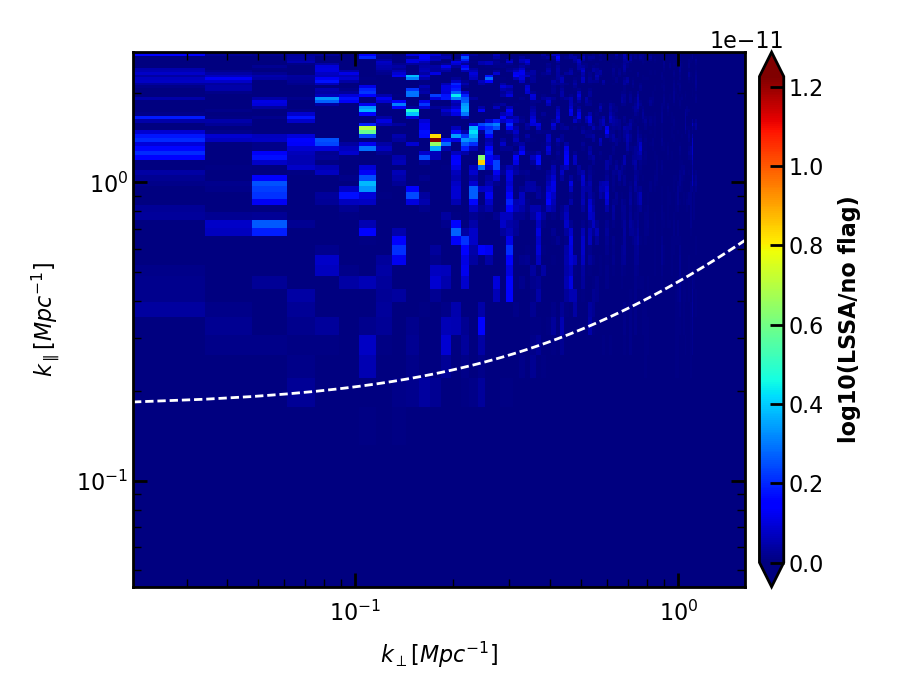} &
       \includegraphics[width=3.5in,height=2.5in]{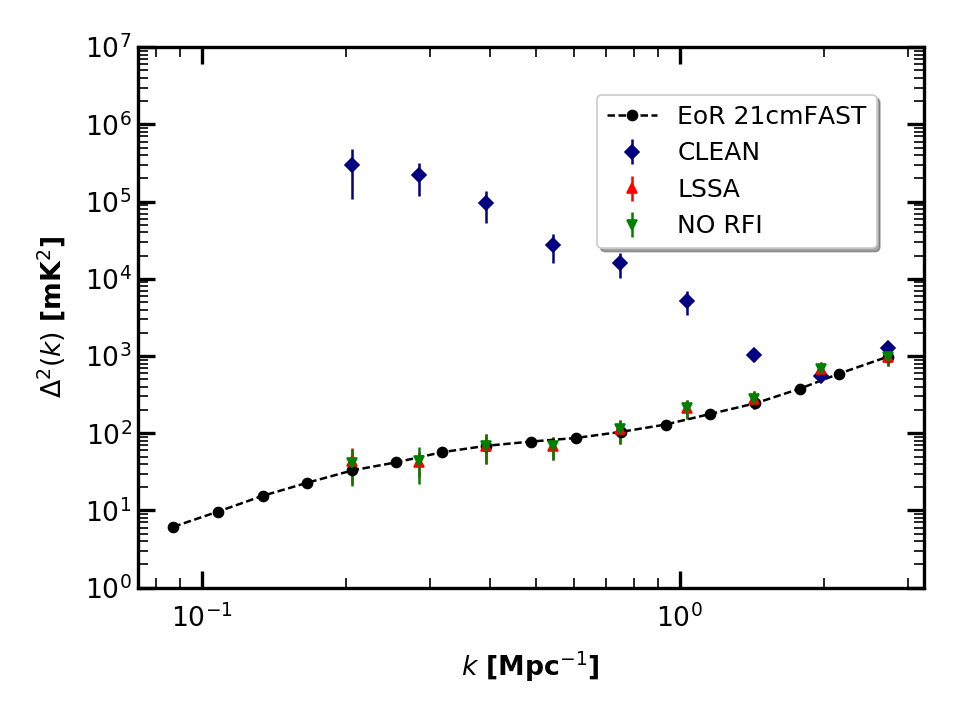}    
      
\end{tabular}      
         
    \caption{ This figure is same as Fig. \ref{mwa_flag_2d}, but for the {\bf Case IV}. Here we introduced ultra-faint broadband RFI flagging in addition to the random and periodic flagging. A 2 MHz chunk at the center of the band is being flagged, which corresponds to broadband RFI (DTV signals). }
    \label{broadband}
\end{figure*}

 {\bf Case III-B}: The results for the  increased flagging channel width (300 kHz) and a decreased  periodicity (in every 0.64 MHz)  than {\bf Case III-A}  are shown in Fig. \ref{mwa_flag_2d_small_period}. The results are nearly similar to the {\bf Case III-A} when we apply BH and LSSA algorithms to the flagged data set. The application of the BH tapering function shows a significant foreground leakage above the horizon. However, applying the LSSA algorithm shows that this leakage is negligible. We also find that, unlike  {\bf Case III-A}, the application of CLEAN gives higher foreground leakage into the EoR window. Note that, here also, we use the tolerance parameter to be $10^{-11}$ in the CLEAN algorithm to reduce the foreground residual. The spherically averaged 3D power spectrum also shows the same feature. We get a large bias due to the foreground leakage at almost all scales when applying the CLEAN algorithm. However, we do not find any such bias in the spherically averaged power spectrum when applying the LSSA algorithm. The analysis of these two cases, {\bf Case III-A and Case III-B}, show that CLEAN is not an optimal way to restrict foreground leakage in the presence of periodic flagging (with any periodicity) along with random flagging, and it gives a  significant bias to the estimated 3D power spectrum. However, the foreground leakage due to the periodic RFI flagging can be restricted with the application of LSSA, and the modes within the EoR window become available again for detection of the cosmological \hi\ power spectrum.\\

 {\bf Case IV}: We show the results for the broadband RFI flagging scenario in  Fig. \ref{broadband}. We find that when a large frequency chunk is missing due to ultra-faint broadband RFI flagging \citep{Wilensky2019PASP..131k4507W}, the CLEAN algorithm shows a significant amount of foreground leakage in comparison with all other cases. The application of the BH tapering function shows similar results as with previous cases. However, applying the LSSA algorithm constrains the leakage during the FT along the frequency axis in this case. The spherically averaged power spectrum also shows the presence of a substantial bias in the estimated power spectrum in the case of  CLEAN. In contrast, we can still recover the Fourier modes inside the EoR window in the case of LSSA. The LSSA can give an unbiased and robust estimate of the underlying power spectrum in the presence of broadband RFI flagging in addition to the random and periodic flag.

The different flagging scenarios used in this comparative analysis prove that LSSA gives a more sensitive and robust way to estimate the cosmological \hi\ power spectrum from the observed data. However, a sufficiently deep CLEANing in the delay space can mitigate the foreground leakage issue in the presence of random RFI flagging only.

\section{Application to the  \texorpdfstring{\lowercase{u}GMRT DATA}{}}
\label{Data}
Here, we extend our comparative analyses to a real observation of the ELAIS-N1 field observed with the uGMRT (in GTAC cycle 32) during May 2017. The details of the observation, a brief discussion on  data analysis and the estimation of  the  power spectrum from the observed visibility data set are discussed in the following sub-sections.

\begin{table}
\caption{Observational details of the target field ELAIS N1 for four observing sessions}
	
\begin{tabular}{ll}
\hline
\hline
Project code & 32\_120 \\
Observation date & 5, 6, 7 May 2017 \\
                 &  27 June 2017\\
\hline
Bandwidth &  200 MHz\\
Frequency range & 300-500 MHz\\
Channels & 8192\\
Integration time & 2s\\
Correlations & RR RL LR LL\\
Total on-source time & $\sim $13 h (ELAIS N1)\\
Working antennas & 26 \\
\hline
Pointing centre & $16^{h}10^{m}01^{s}$  $+54^{d}30^{m}36^{s}$ (ELAIS N1)\\
                
\hline                 
\hline                 
\end{tabular}
\label{observation}	    
\end{table}

\subsection{Observations and Data Analysis}

The data was taken over four nights, where the total integration time, including calibrators, is 25 hours. The total bandwidth (200 MHz) between 300-500 MHz is divided into 8192 channels, which results in a 24 kHz frequency resolution. The raw data has a time resolution of 2 sec. The observational details are tabulated in Table \ref{observation}. The details of data analysis are presented in \citep{ArnabB2019MNRAS.490..243C}. The raw data was flagged for RFI with AOFLAGGER \citep{Offringa2012A&A...539A..95O} and then averaged to 16 sec time resolution to reduce the data volume. About 40\% of the data was flagged after RFI flagging, calibration, and self-calibration steps. We have not averaged the data across frequency coordinate to get the maximum $k_{\parallel}$ modes and work with 24 kHz frequency resolution. This choice will allow us to access a larger region of the EoR window above the foreground wedge. 

\begin{figure*}
    \centering
   
\begin{tabular}{cc}
       \includegraphics[width=3.5in,height=3.in]{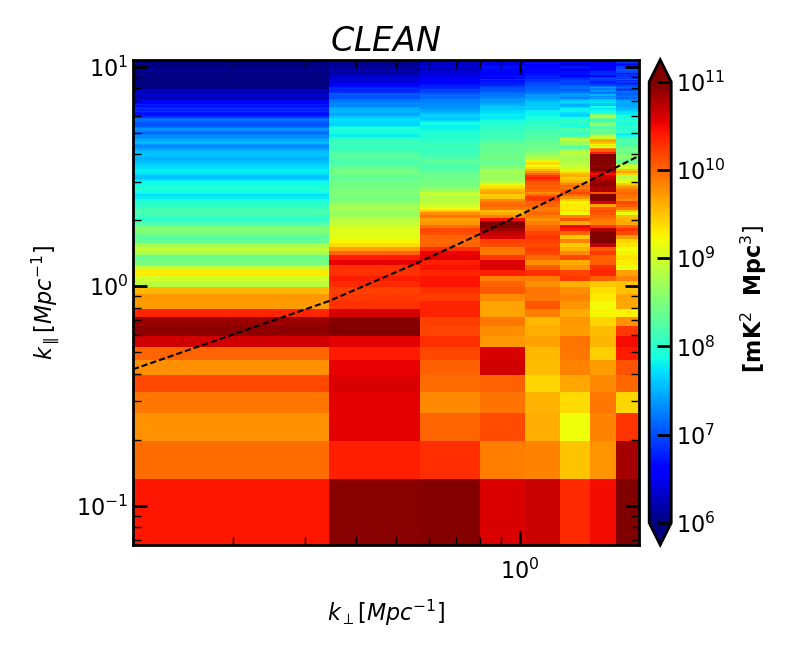} & 
        \includegraphics[width=3.5in,height=3.in]{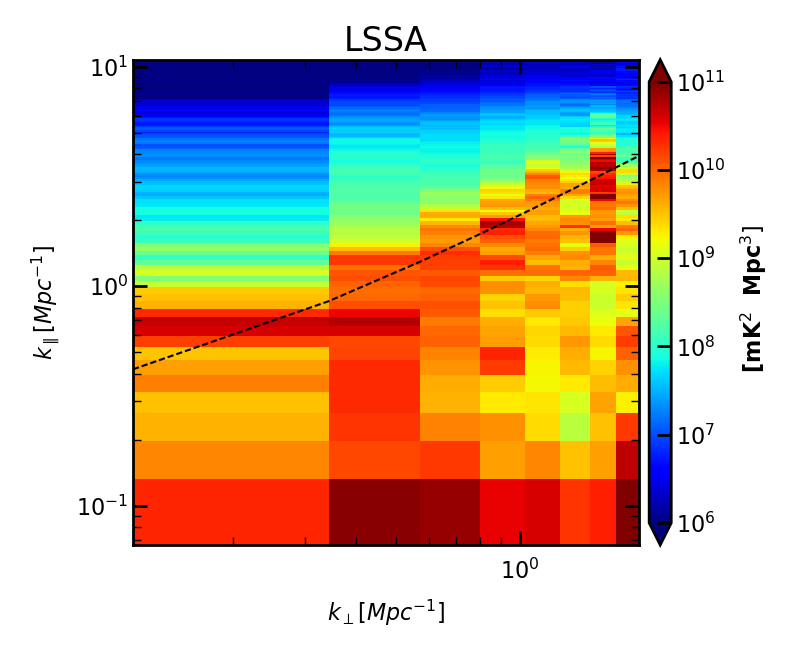} \\ 
     \end{tabular}      
         
    \caption{The cylindrically averaged 2D power spectrum after applying the {\bf CLEAN} (left panel) and LSSA (right panel) to the one night's data set  of the ELAIS N1 field observed with the uGMRT. The data spans a 8 MHz bandwidth around 310 MHz ($z=3.58$). The black dashed line in both panels is the horizon limit. }
    \label{wedge_EN1}
\end{figure*}

\subsection{Power Spectrum estimation from the data}
In this work, we show the result of a small 8 MHz chunk around 304 MHz ($z \sim 3.6$) of one night's data ($\sim$ 6 hours including calibrators). This demonstrates the effect of CLEAN and LSSA algorithms in estimating the power spectrum with realistic flagging applied to the observed visibilities. In our previous work \citet{Arnab2021ApJ...907L...7C}, we have already analyzed the whole data set and put the upper limit on the post-EoR \hi\ 21 cm power spectrum at redshifts $z=1.96,2.19,2.62,3.58$. In \citet{Arnab2021ApJ...907L...7C}, we have used the CLEAN algorithm to mitigate the missing samples issue due to RFI flagging. However, here we also apply the LSSA along with CLEAN on a single night data set to compare the effectiveness of these two algorithms in a realistic scenario. 

The details of the estimation of the power spectrum from the observed uGMRT data set are mentioned in \citet{Arnab2021ApJ...907L...7C} and we briefly mention the methodology here. The simple squaring of visibilities as described in Eqn. \ref{ps_eqn} to estimate the power spectrum will result in positive noise bias since the observed data also contains noise \citep{Somnath2003JApA...24...23B, Ali2008MNRAS.385.2166A}. To avoid this, we grouped the visibilites within a $uv$-cell whose dimension is governed by the inverse of the half-power beamwidth of the primary beam ($\theta_{\mathrm{HPBW}}$). We then cross-correlate the visibilities associated with a particular $uv$-cell. This results in a correlation matrix for each $uv$-cell. The diagonal terms of that matrix are the self-correlation of visibilities and contain the positive noise bias. The off-diagonal terms are the cross-correlation of visibilities that lie inside the $uv$-cell. The noise is expected to be uncorrelated between two visibilities, and so the off-diagonal terms do not contain the positive noise bias \citep{Ali2008MNRAS.385.2166A}. We take the average of the off-diagonal terms of a correlation matrix of a particular $uv$-cell and quote it as an estimated power for that baseline. The underlying assumption is that the cosmological \hi\ signal is correlated among visibilities that fall within a $uv$-cell whose dimension is inverse of the primary beam. This formalism is proposed by  \citet{Somnath2003JApA...24...23B} and investigated in detail using observation by \citet{Ali2008MNRAS.385.2166A}, \citet{Abhik2012MNRAS.426.3295G}, \citet{Samir2016MNRAS.463.4093C}, \citet{somnath2019MNRAS.483.5694B}, \citet{Arnab2021ApJ...907L...7C}, \citet{Srijita2021MNRAS.501.3378P}.

\begin{figure}
    \centering
        \includegraphics[width=3.5in,height=3.in]{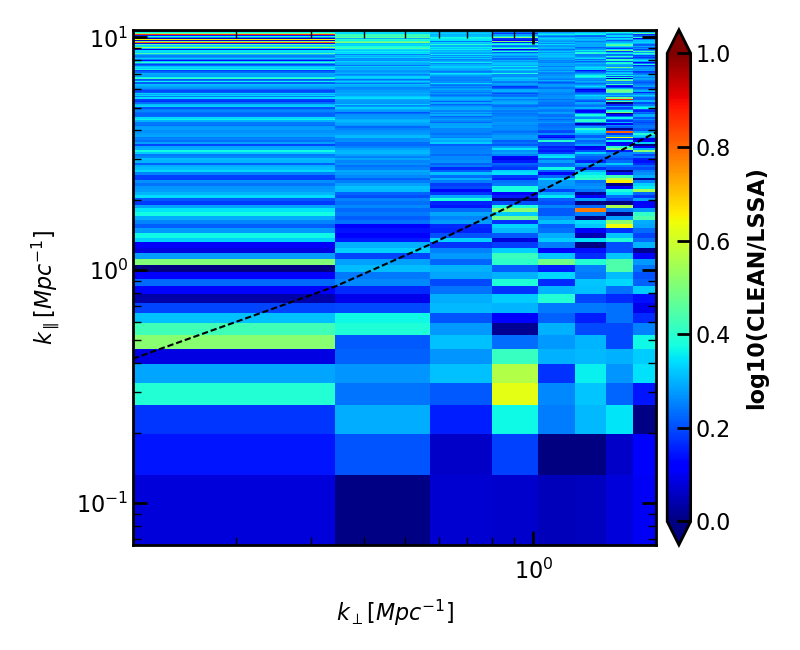} 
    \caption{The ratio of the 2D power spectrum between LSSA and CLEAN algorithms after applying those to the data set of uGMRT. This plot is basically the ratio of the 2D power spectrum shown in Fig. \ref{wedge_EN1}. }
    \label{ratio_EN1}
\end{figure}

We have applied both CLEAN and LSSA to transform the observed visibility data from frequency to delay domain in the presence of realistic flagging and then estimate the power spectrum following the method as described above. In Fig. \ref{wedge_EN1}, we show the cylindrically averaged 2D power spectrum after applying {\it CLEAN} (left panel) and LSSA (right panel) algorithms to the observed visibilities.   We find that both of {\it CLEAN} and LSSA can restrict the foreground leakage above the foreground wedge and give nearly identical results. Above the horizon line, there is a  region where the value of the power spectrum is much less in comparison to the power inside the wedge. This region is relatively free of foreground contamination, and one can spherically average the $\bold k_{\perp}$ and $k_{\parallel}$ modes of that region to estimate the 3D power spectrum. In Fig. \ref{ratio_EN1}, we show the ratio of the 2D power spectrum estimated using CLEAN and LSSA  algorithms. The ratio is close for most of the $k$-modes above the horizon limit. This shows that in the presence of realistic random flagging of the uGMRT data set, both  CLEAN and LSSA give nearly identical results and are consistent with each other.   Note that, here, we do not estimate the spherically averaged power spectrum of cosmological \hi\ signal with this small data set. We only demonstrate the effectiveness of {\it CLEAN} and LSSA  algorithms in case of realistic flagging due to RFI. The 3D spherically averaged power spectrum using the modes free of foreground contamination above the horizon line for different night's data set as well as for the combined data set are already presented in \citet{Arnab2021ApJ...907L...7C}. 

\section{Summary and Conclusion}
\label{Colclusion}
 The flagging of RFI-contaminated channels can cause excess power leaking into the EoR window (above the horizon limit) in the 2D cylindrical power spectrum. The non-uniform sampling of the bandpass due to RFI excision causes large sidelobes in the delay space PSF. The convolution of this non-ideal delay space PSF with the delay spectrum of the data causes foreground leakage into the EoR window. One dimensional CLEAN algorithm in delay space and Least Square Spectral Analysis (LSSA)  are two widely used algorithms, which have been applied to the observed data set to mitigate this foreground leakage issue. In this work, we have done a comparative analysis of these two algorithms using simulations in the presence of different realistic flagging scenarios. 

We have simulated the data for SKA-1 low array configuration with cosmological \hi\ 21 cm signal and foreground signals as sky model at redshift $z \sim 9$. We have compared three different RFI flagging scenarios. At first, we randomly flagged 20\% channels of the simulated data and then applied CLEAN and LSSA to restrict the foreground leakage into the EoR window ({\bf Case II}). We estimate the 2D cylindrically averaged power spectrum and compare the result with the no RFI scenario, i.e., no flagging. We have found that both of these algorithms can restrict the foreground leakage into the EoR window for random flagging of channels. We spherically averaged the modes inside the `EoR window' to estimate the 3D power \hi\ spectrum. We have compared the 3D power spectrum with the no RFI scenario as well as with the injected \hi\ 21 cm signal to our simulation, as taken from 21cmFAST. The results are consistent with each other. We conclude that in the presence of random flagging, the application of CLEAN and LSSA can recover the $k-$modes to extract the cosmological \hi\ 21 cm signal power spectrum.

Then we study the case where we have flagged 200 KHz channels in every 1.28 MHz across the whole 8 MHz band of our simulated data set in addition to 20\%  random flagging ({\bf Case III-A}). This is identical to observed data with the MWA telescope.  We have also changed the periodicity and flagging channel width ({\bf Case III-B}), i.e., we flag 300 KHz channels in every 0.64 MHz across 8 MHz bandwidth. We applied CLEAN and LSSA and compared the 2D and 3D power spectrum with no RFI scenario and with the input \hi\ 21 cm power spectrum. We have found that the CLEAN algorithm fails to restrict the foreground leakage for both cases, and there is an extra bias in the 3D power spectrum. However, the LSSA method can limit the foreground leakage even in the presence of periodic flagging. We can average the modes above the `wedge' and reconstruct the \hi\ 21 cm power spectrum using the LSSA algorithm. We conclude from this analysis that LSSA works better than CLEAN and gives an unbiased estimate of the power spectrum in the presence of periodic flagging.     

 Lastly, we study the scenario where a 2MHz wide frequency chunk has been flagged from the center of the band in addition to the random and periodic flagging. This scenario corresponds to the ultra-faint broadband RFI due to the dish TV signal. After applying different algorithms for this case, we found that CLEAN gives a significant bias in the power spectrum and fails to restrict the foreground leakage. But, LSSA can constrain the leakage and provide an unbiased estimate of the \hi\ power spectrum.

We have also applied these algorithms to the observed data set of the ELAIS N1 field using uGMRT to verify these findings with realistic flagging. However, we only choose a small  8 MHz chunk around 310 MHZ of one night's data (6 hours) and estimate the 2D power spectrum using CLEAN and LSSA algorithms. Here also, we have found that both of these algorithms give identical results in restricting the foreground contamination above the `wedge', since the uGMRT data contains the random flagging only. We can limit the foreground spillover with these algorithms and extract the cosmological \hi\ signal from the uGMRT observation. The first upper limit on post-EoR \hi\ signal using this data set has already been presented in \citet{Arnab2021ApJ...907L...7C}, where we have used the CLEAN algorithm to mitigate the missing sample issue. However, this analysis with the data shows that the power spectrum estimated using the LSSA algorithm is consistent with the CLEAN algorithm.

In summary, this paper explores the effectiveness of 1D CLEAN and LSSA algorithms to deal with missing frequency channel issues in CD/EoR observations. We show that both of these algorithms are effective in the presence of random flagging. However, CLEAN fails in the presence of periodic and broadband RFI flagging. But, the LSSA can still restrict the foreground spillover beyond the horizon limit and give a robust estimate of the \hi\ 21 cm power spectrum. This methodology gives a sensitive estimation of \hi\ power spectrum even in the case where 20\% random channels are flagged, as well as a periodic flagging of channels have been applied across the band. The comparative analysis presented here establishes the advantages and necessity of using one algorithm over the other to estimate \hi\ power spectrum from the observed data infested with RFI. This work is relevant for upcoming telescopes with CD/EoR science goals like the SKA \footnote{\url{https://www.skatelescope.org/}}, HERA \footnote{\url{https://reionization.org/}}  and also for the ongoing projects like the LOFAR \footnote{\url{http://www.lofar.org/astronomy/eor-ksp/lofar-eor-project/lofar-epoch-reionization-project.html}}, MWA \footnote{\url{https://www.mwatelescope.org/science/epoch-of-reionization-eor}}, PAPER \footnote{\url{http://eor.berkeley.edu/}}, etc. However, it should be noted that the missing channels issue is not only limited to CD/EoR frequencies but also to post-EoR redshifted frequencies which are important for the \hi\ 21cm intensity mapping experiments like the CHIME \footnote{\url{https://chime-experiment.ca/en}} , TIANLAI \footnote{\url{https://tianlai.bao.ac.cn}}, HIRAX \footnote{\url{https://hirax.ukzn.ac.za/}}, the SKA-1 mid \footnote{\url{https://www.skatelescope.org/mfaa/}}  and OWFA \footnote{\url{http://www.ncra.tifr.res.in/ncra/ort}}.\\

{\bf ACKNOWLEDGEMENTS}\\
We thank the anonymous referee  and the scientific editor for helpful comments and suggestions that have helped to improve this work. 
AD would like to acknowledge the support of EMR-II under CSIR No. 03(1461)/19. AM would like to thank Indian Institute of Technology Indore for supporting this research with Teaching Assistantship. \\

{\bf DATA AVAILABILITY}

The simulated data cube used in this analysis will be shared on reasonable request to the corresponding author.  The point source catalog of ELAIS-N1 field at 400 MHz compiled from uGMRT observation can be found here \url{https://github.com/Arnabiitindore/Catalog-of-ELAIS-N1-field} and in the link \url{https://doi.org/10.5281/zenodo.5822497}. \\

{\bf $Software$ :}

This work heavily relies on the Python programming language (\url{https://www.python.org/}). The packages used here are pyuvdata (\url{https://github.com/RadioAstronomySoftwareGroup/pyuvdata}; aipy (\url{https://github.com/HERA-Team/aipy/tree/main/aipy}); \citealt{Hazelton2017zndo...1044022H}), astropy (\url{https://www.astropy.org/}; \citealt{Astropy2013A&A...558A..33A,Astropy2018AJ....156..123A}), numpy (\url{https://numpy.org/}) \citealt{van_der_Walt2011CSE....13b..22V,Harris2020Natur.585..357H}, scipy (\url{https://www.scipy.org/}) \citealt{Virtanen2020NatMe..17..261V}, matplotlib (\url{https://matplotlib.org/}) \citealt{Hunter2007CSE.....9...90H}, OSKAR (\url{https://github.com/OxfordSKA/OSKAR}), WSCLEAN(\url{https://gitlab.com/aroffringa/wsclean}).

\bibliographystyle{aasjournal}
\bibliography{clean}

\listofchanges
\end{document}